\documentclass[
reprint,
superscriptaddress,
 amsmath,amssymb,
 aps,
 prx
]{revtex4-2}

\usepackage{graphicx}
\usepackage{dcolumn}
\usepackage{bm}
\usepackage{xcolor}
\usepackage[normalem]{ulem}
\usepackage[english]{babel}
\usepackage{braket}
\usepackage{units}
\usepackage[T1]{fontenc}

\usepackage{ifthen}
\newboolean{showcomments}
\setboolean{showcomments}{true}

\makeatletter
\newcommand{\mynote}[3]{%
  \ifthenelse{\boolean{showcomments}}{%
   \fbox{\bfseries\sffamily\scriptsize#1}%
   {\small$\blacktriangleright$\textsf{\emph{\color{#3}{#2}}}$\blacktriangleleft$}}%
  {%
   \@bsphack
   \@esphack
  }%
}
\makeatother

\newcommand{\Eq}[1]{Eq.\,(\ref{#1})}

\newcommand{\Eqs}[2]{Eqs.\,(\ref{#1}-\ref{#2})}

\newcommand{\Fig}[1]{Fig.\,\ref{#1}}

\usepackage{ulem}
\usepackage{xcolor}
\usepackage{graphicx}
\usepackage{dcolumn}
\usepackage{bm}
\usepackage[colorlinks]{hyperref}

\newcommand{\avg}[1]{\langle{#1}\rangle}

\newcommand{\expect}[1]{\mathbb{E}\bigl[ {#1}\bigr]}

\begin{document}

\preprint{APS/123-QED}

\title{Single-shot Quantum State Classification via Nonlinear Quantum Amplification}

\author{Elif Cüce}
\affiliation{Department of Electrical and Computer Engineering, Princeton University, Princeton, NJ 08544, USA}
\author{Saeed A. Khan}
\affiliation{School of Applied and Engineering Physics, Cornell University, Ithaca, NY 14853, USA}
\author{Boris Mesits}
\affiliation{Department of Physics and Astronomy, University of Pittsburgh, Pittsburgh, PA 15260, USA}
\affiliation{Department of Applied Physics, Yale University, New Haven, CT 06511, USA}
\affiliation{Yale Quantum Institute, Yale University, New Haven, Connecticut 06520, USA}
\author{Michael Hatridge}
\affiliation{Department of Applied Physics, Yale University, New Haven, CT 06511, USA}
\affiliation{Yale Quantum Institute, Yale University, New Haven, Connecticut 06520, USA}
\author{Hakan E. Türeci}
\affiliation{Department of Electrical and Computer Engineering, Princeton University, Princeton, NJ 08544, USA}
\date{\today}

\begin{abstract}
Quantum amplifiers are intrinsically nonlinear systems whose performance limits are set by quantum mechanics. In quantum measurement, amplifier operation is conventionally optimized in the linear regime by maximizing signal-to-noise ratio, an objective that is well-suited to parameter estimation but is typically insufficient for more general tasks such as arbitrary quantum state discrimination. Here we show that single-shot quantum state classification can benefit from operating a quantum amplifier outside the linear regime, when the measurement chain is optimized end-to-end for a task-specific cost function. We analyze a realistic superconducting readout architecture that includes state preparation, cryogenic nonlinear amplification, and room-temperature detection with finite noise. By introducing performance metrics tailored to state discrimination, we identify operating regimes in which nonlinear amplification provides a measurable advantage and clarify the trade-offs that ultimately limit classification fidelity. Building on these results, we propose a  qubit readout architecture without cavity displacement that exploits nonlinear amplification to enhance single-shot state discrimination performance. Our results establish the practical value of nonlinear quantum amplifiers for quantum state discrimination and lay the foundation for a broader program to develop a general, end-to-end framework for resource-constrained optimization of nonlinear amplification in quantum information processing tasks.
\end{abstract}

\maketitle

\section{\label{sec:intro}Introduction}

\begin{figure*}
\includegraphics[width=\linewidth]{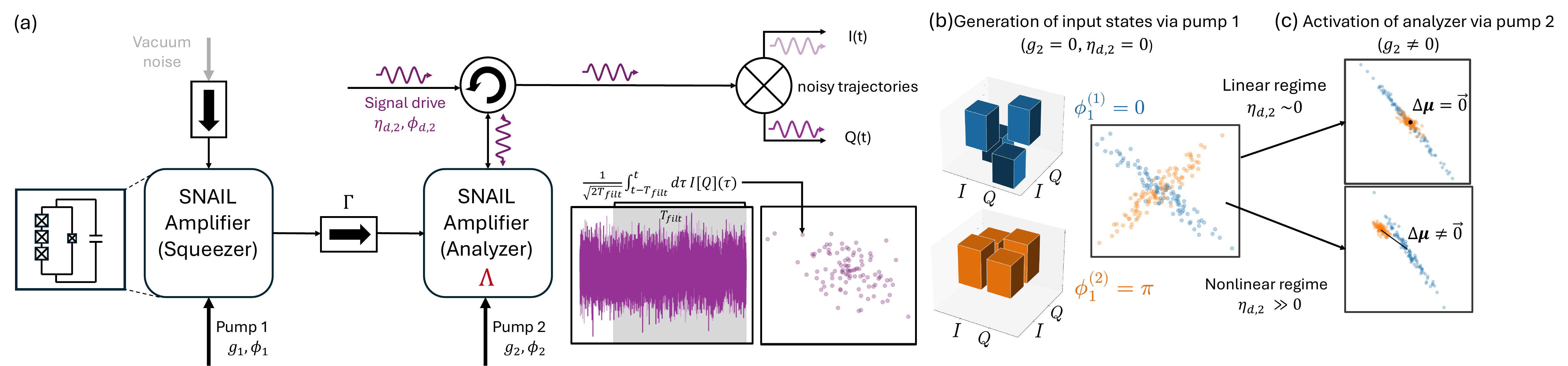}
\caption{\label{fig:intro}(a) The composite two-SNAIL system for state generation and discrimination. The pump applied to the \textit{squeezer} (pump 1) sets which input state is generated, (1-blue) or (2-orange). Two additional drives at the pump frequency $\omega_p$ (pump 2) and at $\omega_p/2$ (signal drive with strength $\eta_{d,2}$) act on the \textit{analyzer} and impact the overall classification performance. The two SNAIL amplifiers are identical except that the analyzer has a non-zero Kerr nonlinearity ($\Lambda$) (b) Intracavity quadrature covariance matrix elements and quadrature measurements for two input states with the analyzer pump off ($g_2 = 0$). (c) When the analyzer is on, the output-state means coincide for linear operation ($\eta_{d,2} \approx 0$) and the classification fidelity is 0.84, but are separated for nonlinear operation with a sufficiently-large signal drive ($\eta_{d,2} \gg 0$) and the classification fidelity is 1 (this is the same operating point as in Fig.~2(d-2)).}
\end{figure*}

Rapid, high-fidelity quantum nondemolition measurement requires amplification of an electromagnetic probe field near the limits set by quantum mechanics. In superconducting-qubit readout, where the probe lies in the microwave domain and the qubits are only weakly nonlinear, this is achieved by pre-amplifying the outgoing field with a cryogenic amplifier at base temperature, prior to further amplification and digital processing at room temperature.~\cite{blais_circuit_2021}. Although many cryogenic amplifiers are intrinsically nonlinear, they are conventionally operated in their linear regime, where the relevant quantum limits are well understood~\cite{haus_quantum_1962, caves_quantum_1982, clerk_introduction_2010, aumentado_superconducting_2020, khan_practical_2024}. Can operating a cryogenic quantum amplifier in a nonlinear regime provide a fundamental performance advantage beyond what is achievable in linear, quantum-limited operation? We address this question through a hardware-realistic analysis of a class of quantum state–discrimination tasks in which operating an experimentally available parametric oscillator in a genuinely nonlinear regime is optimal. In contrast to the prevailing emphasis on linear, Kerr-mitigated amplification~\cite{frattini_optimizing_2018,sivak_kerr-free_2019}, we identify operating regimes where nonlinear amplification yields a measurable classification advantage and quantify how nonlinear distortion, effective noise, and downstream added noise set the ultimate limits on classification fidelity.

In classical signal processing, nonlinear amplification has long been used as a deliberate functional element for tasks where linear gain is suboptimal. The underlying principle is that when the objective is not faithful waveform reproduction but rather feature extraction, detection, or classification, nonlinear transformations can dramatically improve performance under resource constraints~\cite{Kay_Fundamentals_1998, VanTrees_Detection_1968}. By comparison, the deliberate use of nonlinear amplification for processing quantum-domain signals has received relatively little attention to date. Ref.~\cite{bondurant_quantum_1993} examined an optical amplifier based on a non-linear interferometric setup. In microwave quantum circuits, non-linear dynamical transduction either provided by a resonant cavity~\cite{reed_high-fidelity_2010} or by a Josephson junction tuned close to its `bifurcation point'~\cite{vijay_invited_2009} has been explored to attain high-sensitivity and speed in qubit readout. A comprehensive analysis of the quantum limits of a broad class of nonlinear amplifiers was recently presented in Ref.~\cite{epstein_quantum_2021}, and stands out in the contemporary literature. 

When the objective is rapid, high-fidelity measurement for quantum state discrimination rather than an analog task, quantum limits on amplification alone are not sufficient to determine performance. Additional considerations, including amplifier nonlinearity, demodulation strategy, integration time, and the noise incurred upon extraction of the signal to room temperature, play a critical role~\cite{gambetta_protocols_2007,silveri_theory_2016,clerk_introduction_2010,khan_practical_2024}.  Moreover, through a concrete example, we show that the optimal strategy depends explicitly on the measurement objective and, more generally, on the specific cost function being maximized.

The purpose of this article is to present a systematic study of the quantum-limited operating regimes of a specific class of nonlinear amplifiers, with an eye on end-to-end resource-aware optimization of a task-relevant cost-function. We highlight the central role of task-specific optimization by focusing on a quantum state discrimination problem that is poorly served by amplifiers optimized for linear, quantum-limited operation followed by linear detection and post-processing. Our analysis identifies the operating regimes in which nonlinear amplification yields a measurable performance advantage and clarifies the trade-offs that ultimately set the limits on achievable measurement fidelity.

Our work is motivated by recent proposals in quantum computational sensing~\cite{khan_quantum_2025}, which articulate a framework for formulating quantum measurement and sensing tasks in close analogy with conventional computational sensing and imaging approaches. This line of work takes the perspective that the detection of weak signals is often performed with the ultimate aim of performing a specific computation on the signal, in service of completing a given task. Consequently, weak-signal detection or sensing can be combined with processing by a suitably-matched quantum mechanical system prior to detection. This line of research fuses concepts from quantum sensing and quantum machine learning, and has demonstrated that the resulting quantum computational sensors can provide advantages for tasks such as classification~\cite{zhuang_physical-layer_2019, sinanan-singh_single-shot_2024, chin2024quantumentanglementenablessingleshot, allen2025quantumcomputingenhancedsensing} and function approximation~\cite{eldredge2018optimal, khan_quantum_2025} in comparison to sensing schemes alone. For example, within this framework, circuit-based quantum computational sensors have been shown to approach the fundamental performance bounds imposed by quantum mechanics for a broad class of sensing  tasks~\cite{ilias_end--end_2025}. 

The present work lies in the same domain, but differs in scope and methodology. Firstly, rather than adopting a gate-based and lossless quantum computational sensor, we consider a continuous-variable, dynamical, driven-dissipative sensor. Prior work has shown that nonlinear resources available in such cryogenic sensors can be used for certain quantum information processing applications~\cite{khan_neural_2025}. Here, we consider a sensor that is closely-modeled on practical quantum amplifiers, and therefore features amplification and dissipation as key components. Using a model-based approach in which the full measurement chain, including signal generation, propagation, amplification, and detection, is described numerically, we assess performance and identify optimal operating regimes for such a device for quantum state discrimination tasks under constrained resources.

We next examine the optimal operating conditions for nonlinear amplification in dispersive qubit readout without cavity displacement, and present a detailed analysis of an architecture incorporating these findings. We show that the optimal operating point is architecture-dependent and, in this setting, lies in the regime $\chi < \kappa/2$, where $\chi$ is the qubit-induced dispersive shift and $\kappa$ is the resonator linewidth. This stands in contrast to conventional dispersive readout with linear, phase-preserving amplification, where maximizing measurement rate and SNR typically favors operation in or near the strong-dispersive regime, $\chi = \kappa/2$, for fixed photon number and coherence constraints.

The rest of this paper is organized as follows. In Sec.~\ref{sec:syst-desc-class}, we describe the mathematical model of the composite system that describes both the generation of states to be discriminated and the measurement chain including the nonlinear amplifier. In Sec.~\ref{sec:analyzer-opt}, we define metrics to quantify the classification performance and optimize the externally tunable parameters of the system. Next, in Sec.~\ref{sec:th}, we analyze the effect of the noise from the electronic amplification on the performance metrics. Finally, in Sec.~\ref{sec:qr}, we propose a scheme to perform dispersive qubit readout in a regime without cavity displacement, and find the optimal encoding parameters for this scheme. In Sec.~\ref{sec:conc}, we summarize our conclusions and outline directions for future work.  

\section{\label{sec:syst-desc-class}Description of the composite system and the classification task}

We consider the task of discriminating two states of a resonator by amplifying the radiation that leaks out of the resonator with an amplifier that intrinsically is nonlinear. To this end, we introduce a suitable objective function of the single-shot sampled output of the amplifier that we require to be maximized. We consider maximization with respect to parameters of the amplifier that can be parametrically controlled during the experiment.  

For a single shot, there is a fundamental physical limit to the expected maximum of the objective function that is set by the quantum theory, regardless of whether the amplifier is operated in the linear or the nonlinear regime. We are interested in the optimal achievable objective function within a given class of amplifiers. We note that the achievable maximum heavily depends on the objective function itself and the states to be discriminated. In the case of dispersive qubit readout, the task can be framed as a quantum state discrimination task where two distinct pointer states of a readout resonator are generated conditional on the two possible states of the qubit. Here the physical limits are well understood when the downstream amplifier and the subsequent detector is linear~\cite{haus_quantum_1962,caves_quantum_1982,gambetta_protocols_2007,clerk_introduction_2010,silveri_theory_2016,aumentado_superconducting_2020}.

In this work, we address a more challenging problem in which the two pointer states are resonator states with identical mean amplitudes and are distinguished solely by their second moments. We do not specify a particular physical mechanism for generating these conditional pointer states, and focus on a regime in which quantum back-action on the qubit, outside the scope of the present study, is expected to be minimal. Specifically, we consider the limit in which the mean fields of the pointer states vanish, so that the influence of the qubit can be neglected and the problem reduces to the discrimination of two zero-mean resonator states differing only in their fluctuations. Earlier work~\cite{epstein_quantum_2021} has shown that an ideal nonlinear amplifier in principle allows one to measure any normal operator (of its input) with a linear detector while adding a half-quantum of vacuum fluctuations as noise at the output, but what types of measurements can be enabled by common experimentally realizable nonlinearities was not analyzed. Here, for the first time to our knowledge, we consider the end-to-end optimization of an experimentally realizable superconducting readout chain, including state preparation, directional coupling to a cryogenic nonlinear amplifier, and the noise floor of a subsequent room-temperature detection stage. This optimization requires the quantum dynamical modeling of the entire measurement chain and the sampling of measurement-conditioned trajectories observed at the linear detector, which we analyze using a recently developed cumulant expansion technique~\cite{khan_neural_2025,khan_physical_2021}. We derive analytical performance metrics that guide our understanding of the behavior of the amplifier in certain well-understood operational limits. We then analyze the performance across an experimentally relevant parameter space and identify optimal, in situ tunable operating points. Finally, we provide intuitive physical reasoning for these optimal choices, highlighting the properties that underlie the classification mechanism. 

A schematic of the complete measurement chain, including preparation of the two pointer states and the nonlinear amplification stage, is shown in \Fig{fig:intro}. The setup employs two Superconducting Nonlinear Asymmetric Inductive eLement (SNAIL) oscillators, which serve as the fundamental building blocks of state-of-the-art cryogenic microwave amplifiers~\cite{frattini_optimizing_2018}. The first SNAIL oscillator, referred to as the \textit{squeezer}, is used to generate two distinct pointer states, which are squeezed vacuum states with orthogonal squeezing directions. These states are subsequently discriminated by the second SNAIL oscillator, referred to as the \textit{analyzer}. Both SNAILs are identical in that they share the same bare resonance frequency and the same third- and fourth-order (Kerr) nonlinearity strengths ($g_3$ and $g_4$ in \Eq{eq-H-SNAIL}). The difference between the two lies in their operating conditions. The \textit{analyzer} SNAIL is driven coherently at the signal port, which displaces the mode and leads to an effective contribution of the Kerr term to its dynamics. In contrast, the \textit{squeezer} SNAIL is not subject to a coherent signal drive, and the Kerr term does not contribute appreciably to its dynamics. We therefore neglect the Kerr term for the \textit{squeezer} SNAIL to simplify the model. We note that this set up has a simple linear limit for the \textit{analyzer}, achieved by either turning off the additional coherent drive or setting its nonlinearity $\Lambda=0$, when the analyzer becomes a phase-sensitive linear amplifier. In our numerical analysis, we model all essential elements of a typical superconducting measurement chain, including directional propagation between the \textit{squeezer} and \textit{analyzer} SNAILs, the demodulation at the mixer and the integration of the down-converted measurement trajectory at the room temperature. 

\subsection{\label{sec:syst-desc} The Model}

We describe the measurement-conditioned evolution of the system with the following stochastic quantum master equation (SQME), assuming that the \textit{analyzer} mode undergoes continuous linear measurement:
\begin{equation}
    \dot{\hat{\rho}}^c = \mathcal{L}_{}\hat{\rho}^c + \mathcal{S}_{an}\hat{\rho}^c.
    \label{eq-qsme}
\end{equation}
The Liouvillian of the system, which we describe in detail in this section, is derived from a general model of two SNAILs, representing the \textit{squeezer} and the \textit{analyzer}, that are unidirectionally coupled. The Hamiltonian of a single SNAIL oscillator representing either the \textit{squeezer} or the \textit{analyzer}, with resonance frequency $\omega_s$, incorporating both third- and fourth-order nonlinearities and driven simultaneously with two tones at frequencies $\omega_p$ and $\omega_p/2 = \omega_s$, is given by
\begin{equation}
\begin{split}
    \hat{\mathcal{H}}_{SNAIL}& = \omega_s \hat{a}^{\dagger}\hat{a}  + g_3 \left(\hat{a}+\hat{a}^{\dagger}\right)^3 + g_4 \left(\hat{a}+\hat{a}^{\dagger}\right)^4 
    \\
    &+ \sqrt{\kappa_s}\epsilon_p \left(e^{-i\phi_p}e^{-i\omega_p t} + e^{i\phi_p}e^{i\omega_p t}\right) \left(\hat{a}+\hat{a}^{\dagger}\right) \\
    &+ \sqrt{\kappa_s}\eta_{sig} \left(e^{-i\phi_{sig}}e^{-i\frac{\omega_p}{2} t} + e^{i\phi_{sig}}e^{i\frac{\omega_p}{2} t}\right) \left(\hat{a}+\hat{a}^{\dagger}\right).
    \end{split}
    \label{eq-H-SNAIL}
\end{equation}
Here, the first line includes the bare resonance term, third- and fourth-order nonlinearities of the SNAIL that are non-zero. The second and third lines represent the pump drive and signal drive, respectively. We assume that the \textit{squeezer} has zero Kerr nonlinearity ($g_4 = 0$) and is not driven at resonance frequency ($\eta_{sig}=0$). Due to the presence of the pump and the signal drive on the SNAIL, there will be non-zero population at $\omega_p$ (pump mode) and $\omega_p/2$ (resonant mode). To account for both modes, we introduce the following multimode field ansatz:
\begin{align}
    \hat{a} = \hat{s}e^{i\frac{\omega_p}{2}t} + \hat{p}e^{i\omega_pt},
\end{align}
where $\hat{s}$ represents the signal mode and $\hat{p}$ represents the pump mode. As long as the dynamics of the new operators have bandwidth much smaller than $\omega_p/2$, it is a good approximation to assume $\hat{s}$ and $\hat{p}$ represent independent dynamical degrees of freedom \cite{roy_introduction_2016,roy_quantum-limited_2018}. After plugging in this ansatz and assuming the pump is stiff, we end up with the Liouvillian that we introduce in this section. More details of this derivation is in Appendix~\ref{app-full-derivation}. The Liouvillian $\mathcal{L}$ of the system is composed of three components pertaining to the \textit{squeezer}, nonreciprocal coupler and the \textit{analyzer}:
\begin{equation}
    \mathcal{L} = \mathcal{L}_{sq} + \mathcal{L}_{an} + \mathcal{L}_{coup}.
\end{equation}
$\mathcal{L}_{sq}$ describes the \textit{squeezer},
\begin{equation}
    \mathcal{L}_{sq} \hat{\rho}^c = -i\left[\hat{\mathcal{H}}_{sq},\hat{\rho}^c \right]  +\kappa_1\mathcal{D}[\hat{s}_1]\hat{\rho}^c,
\end{equation}
with
\begin{equation}
    \hat{\mathcal{H}}_{sq} = -\Delta_1 \hat{s}_1^{\dagger} \hat{s}_1 + \frac{g_1}{2}\left(e^{-i\phi_1} \hat{s}_1 \hat{s}_1 + h.c.\right).
\end{equation}
The Hamiltonian for the \textit{squeezer} is expressed in a frame rotating at $\omega_p/2$, where $\omega_p$ is the pump tone frequency. This is the frame in which the squeezing terms are resonant. We assume the system is being pumped on resonance for amplification ($\omega_{1}=\omega_{2}= \omega_p/2$), which makes $\Delta_1 = 0$.  We assume that the pump is stiff and characterized with real-valued constants $g_1$ and $\phi_1$ for the strength and phase respectively of this stiff pump. We verify the validity of this approximation within the operating regime considered in the paper by comparing it to simulations performed without the stiff-pump approximation. Further, $\mathcal{L}_{an}$ describes the \textit{analyzer}, 
\begin{equation}
    \mathcal{L}_{an} \hat{\rho}^c = -i\left[\hat{\mathcal{H}}_{an},\rho^c \right]  +\kappa_2\mathcal{D}[\hat{s}_2]\rho^c,
\end{equation}
with
\begin{eqnarray}
    \hat{\mathcal{H}}_{an} = &&-\Delta_2 \hat{s}_2^{\dagger} \hat{s}_2 + \frac{g_2}{2}\left(e^{-i\phi_2} \hat{s}_2 \hat{s}_2 + h.c.\right) -\frac{\Lambda}{2} \hat{s}_2^{\dagger^{2}}\hat{s}_2^{2}\nonumber\\
    &&+\sqrt{\kappa_2}\eta_{d,2}(ie^{-i\phi_{d,2}}\hat{s}_2e^{-i(\omega_{d,2}-\omega_p/2) t} + h.c).
    \label{eq-an-H}
\end{eqnarray}
The Hamiltonian for the \textit{analyzer} is also expressed in the frame rotating at $\omega_p/2$ for the same reason. With this choice and the assumption $\omega_{2} = \omega_p/2$, $\Delta_2$ only includes the cross-Kerr shift due to the pump. We present this calculation in Appendix~\ref{app-full-derivation}. Again, we assume that the \textit{analyzer} pump is stiff, and is characterized by constants $g_2$ and $\phi_2$ for its strength and phase. Different from the \textit{squeezer}, $\hat{\mathcal{H}}_{an}$ contains the term with coefficient $\Lambda$ that corresponds to the intrinsic Kerr nonlinearity of the SNAIL device. In the context of parametric amplifiers this fourth-order nonlinearity generally acts as a parasitic effect that limits dynamic range and leads to gain compression and unwanted frequency shifts, and is therefore typically treated as an imperfection to be minimized in linear-amplification regimes in order to improve dynamic range and approach quantum-limited performance \cite{boutin_effect_2017, frattini_optimizing_2018, sivak_kerr-free_2019}. In Appendix~\ref{app-full-derivation} we present a derivation of this effective SQME starting with the two-SNAIL Hamiltonian, the conditions under which this model accurately describes the dynamics of the system, as well as the relationship of the model parameters to the original physical parameters of the SNAILs. The final term represents the signal drive on the \textit{analyzer}. We choose a resonant drive on the analyzer mode $\omega_{d,2} = \omega_2 = \omega_p/2$.

The coupling interaction with rate $\Gamma$ is described by the coupling Liouvillian $\mathcal{L}_{coup}$,
\begin{equation}
    \mathcal{L}_{coup} \hat{\rho}^c =-i\left[ \left(i\frac{\Gamma}{2}\hat{s}_1^{\dagger}\hat{s}_2 + h.c. \right),\hat{\rho}^c \right] +\Gamma  \mathcal{D}\left[\hat{s}_1+\hat{s}_2 \right]\hat{\rho}^c.
    \label{eqn-coup-L}
\end{equation}
This term balances a coherent hopping interaction with a dissipative hopping interaction to ensure nonreciprocal transmission from the \textit{squeezer} mode $\hat{s}_1$ to the \textit{analyzer} mode $\hat{s}_2$ only~\cite{metelmann_nonreciprocal_2015,khan_neural_2025}.

Lastly, the measurement operator models a weak measurement of only the \textit{analyzer} mode ($\hat{s}_2$) as described by the measurement operator $\mathcal{S}_{an}$~\cite{wiseman_quantum_2009}:
\begin{eqnarray}
    &\mathcal{S}_{an} \hat{\rho}^c = \sqrt{\frac{\kappa_2}{2}} \left(\hat{s}_2\hat{\rho}^c + \hat{\rho}^c \hat{s}_2^{\dagger} - \left\langle \hat{s}_2 + \hat{s}_2^{\dagger}\right\rangle\right)\frac{dW_I}{dt} \nonumber \\
    &+ \sqrt{\frac{\kappa_s}{2}} \left(-i\hat{s}_2\hat{\rho}^c +i \hat{\rho}^c \hat{s}_2^{\dagger} -i \left\langle \hat{s}_2 - \hat{s}_2^{\dagger}\right\rangle\right)\frac{dW_Q}{dt}.
    \label{eqn-an-meas}
\end{eqnarray}
Here, $dW_{I,Q} = dW_{I,Q}(t)$ are independent Wiener increments, that describe the backaction on the system dynamics conditioned on the homodyne I-Q quadrature measurements of the \textit{analyzer} mode. We include the measurement terms in the description of the system since we want to account for finite measurement resources available in an experimental setting such as a finite integration time in a single run of the experiment. 

\subsection{\label{sec:syst-dyn} System dynamics and measurement}

The numerical solution of \Eq{eq-qsme} necessitates a truncation scheme that remains accurate across widely disparate oscillator occupations within the measurement chain. In the regime considered here, a Fock-space truncation is impractical due to resource constraints, as the \textit{analyzer} can reach large occupations for certain parameter regimes analyzed. Here we employ a cumulant-based approach, which we have extensively benchmarked and found to remain accurate over the range of nonlinearities and dissipation strengths relevant to this work. Cumulant- (or cluster-) based moment closures are a standard tool in quantum optics for reducing the infinite hierarchy of operator moments to a finite, tractable set by discarding connected correlators above a chosen order~\cite{schack_moment_1990}. When higher-order connected correlations remain parametrically small e.g., in near-Gaussian, weakly correlated, or mean-field–like regimes, this truncation provides a controlled and numerically stable approximation that has been extensively benchmarked across representative light–matter and nonlinear optical settings~\cite{schack_moment_1990}. Building on this foundation, we employ a truncated-cumulant scheme recently adapted to efficiently sample stochastic measurement trajectories produced by nonlinear measurement chains in Refs.~\cite{khan_physical_2021,khan_neural_2025}. 

In this section, we present the equations of motion for the lowest-order operators to illustrate the approach and to discuss important details of the numerical pipeline used to generate the $I$-$Q$ plots (such as e.g. \Fig{fig:intro}(b,c)) employed throughout the analysis. For details, see Appendix \ref{app-cumulants}.

The equation of motion of the \textit{analyzer} mode expectation value conditioned on the measurement $\avg{\hat{s}_2}^c$ is given by:
\begin{equation}
    \begin{split}
    d\avg{\hat{s}_2}^c = & dt \Bigg\{ -{\Gamma_1}\avg{\hat{s}_1}^c + \left(i\Delta_2 -\frac{\kappa_2+\Gamma_1}{2}\right) \avg{\hat{s}_2}^c 
    \\
    &-ig_2e^{i\phi_2} \avg{\hat{s}_2^{\dagger}}^c -\sqrt{\kappa_2}\eta_{d,2}e^{i\phi_{d,2}}
    \\
     &+i\Lambda\left[2 C^c_{s_2^{\dagger}s_2} \avg{\hat{s}_2}^c + \avg{\hat{s}_2^{\dagger}}^c\left( \avg{\hat{s}_2}^{c^2} +  C^c_{s_2s_2}\right) \right]  \Bigg\} 
    \\
    &+ \sqrt{\frac{\kappa_2}{2}}\left(   C^c_{s_2s_2} +  C^c_{s_2^{\dagger}s_2} \right)dW_I 
    \\
    &-i \sqrt{\frac{\kappa_2}{2}}\left(C^c_{s_2s_2} -  C^c_{s_2^{\dagger}s_2} \right)dW_Q,
    \end{split}
    \label{eq-s2-STEOM}
\end{equation}
where $C^c_{o_1o_2} = \avg{\hat{o}_1\hat{o}_2}^c - \avg{\hat{o}_1}^{c}\avg{\hat{o}_2}^{c}$ are the conditional second-order cumulants and $\avg{\cdot}^c$ represents the conditional average, due to the presence of the measurement operator in the SQME.

The Heisenberg equations of motion, a generally infinite hierarchy of \textit{linear} equations expressed in moments, when reorganized in cumulants, becomes a set of coupled nonlinear differential equations. The non-zero Kerr nonlinearity of the \textit{analyzer} mode manifests itself in the appearance of nonlinear deterministic terms at lowest order. If $\Lambda=0$, noise-averaged first order equations decouple from equations of motion of second-order cumulants (see Appendix~\ref{app-cumulants} \Eqs{eq-app-teom-s1}{eq-app-teom-s2D}). Note however that the conditional evolution has noise terms that couple to second order cumulants and generate what would be considered multiplicative noise (see Appendix~\ref{app-cumulants} Eqs.~\ref{eq-app-steoms}). We will show that the residual nonlinearity of the amplifier, as captured here, enables the transduction of information encoded in higher-order features of the measured quantum system, specifically the information contained in the second-order correlations of the two pointer states of the \textit{squeezer}. The versatility of STEOM is critical as we are interested in a regime where the \textit{analyzer} mode is sufficiently populated for the Kerr nonlinearity to be effective.

Once the complete set of nonlinear STEOMs are solved to a certain order, the $I$-$Q$ plots typically measured in an experiment are obtained after demodulation at $\omega_p/2$ and integration with a box-car filter. First step in this procedure is the generation of quantum trajectories of the outgoing \textit{analyzer} field which represent the noisy measurement acquired after demodulation:
\begin{subequations}
\label{eqn-noisy-I-Qtraj}
\begin{equation}
\mathcal{I}(t) = \xi_{I}(t) + \sqrt{\frac{\kappa_2}{2}}\left\langle \hat{s}_2 + \hat{s}_2^{\dagger}\right\rangle^c +\sqrt{\Bar{n}_{cl}}\xi_{I}^{cl}(t),
    \label{eqn-noisy_Itrace}
\end{equation}
\begin{equation}
\mathcal{Q}(t) = \xi_{Q}(t) -i \sqrt{\frac{\kappa_2}{2}}\left\langle \hat{s}_2 - \hat{s}_2^{\dagger}\right\rangle^c +\sqrt{\Bar{n}_{cl}}\xi_{Q}^{cl}(t),
    \label{eqn-noisy_Qtrace}
\end{equation}
\end{subequations}
where $\xi_{I,Q}(t) = \frac{dW_{I,Q}}{dt}$, with $dW_{I,Q}$ as in Eq.~(\ref{eqn-an-meas}), and $\xi_{I,Q}^{cl}(t)$ are independent white noise processes modelling electronic amplification noise with strength $\Bar{n}_{cl}$. Moreover, we define intracavity quadrature variables $\hat{I}_2 = \left(\hat{s}_2+\hat{s}_2^{\dagger}\right)/\sqrt{2}$ and $\hat{Q}_2 = -i\left(\hat{s}_2-\hat{s}_2^{\dagger}\right)/\sqrt{2}$ for later use. 

The noisy trajectories in Eq.~(\ref{eqn-noisy-I-Qtraj}) are generally filtered over a filtering period $\mathcal{T}$, to obtain single-shot measurement results: 
\begin{subequations}
    \label{eqn-filt-I-Q}
    \begin{equation}
        I = \frac{1}{\sqrt{2\mathcal{T}}}\int_{t-\mathcal{T}}^{t}d\tau \mathcal{I}(\tau),
        \label{eqn-filt-I}
    \end{equation}
    \begin{equation}
        Q = \frac{1}{\sqrt{2\mathcal{T}}}\int_{t-\mathcal{T}}^{t}d\tau \mathcal{Q}(\tau).
        \label{eqn-filt-Q}
    \end{equation}
\end{subequations}
The scaling with $\mathcal{T}$ is chosen here to keep variance independent of $\mathcal{T}$ in the long filtering time limit, up to corrections that scale as $\mathcal{O}(\mathcal{T}^{-1})$, whereas the mean value of the filtered quadrature scales with $\sqrt{\mathcal{T}}$. See Appendix~\ref{app-filt} for more details on the filtered single-shot measurement results.

\subsection{\label{sec:squeezer-task}Generation of squeezed vacuum states for the example task}

The \textit{squeezer} generates the squeezed vacuum states to be classified. Its output state is controlled by the pump tone: the pump amplitude sets the degree of squeezing, while the pump phase $\phi_1$ fixes the squeezing orientation in phase space (see \Fig{fig:intro}). These properties can be determined analytically by deactivating the \textit{analyzer} in the model, i.e., by setting $g_2 = 0$, $\eta_{d,2}=0$ and $\Lambda = 0$, so that the measured output of the \textit{analyzer} is a proxy for the measured statistical distribution of the \textit{squeezer} state. In this limit the entire system is linear, allowing an exact derivation of the quadrature covariance matrix that is measured at the output of the \textit{analyzer}. The squeezing axis in the measured $I$–$Q$ plane can be shown to be given by $\phi_{\mathrm{sq}} = \frac{\pi}{4} - \frac{\phi_1}{2}$, where the $I$ quadrature defines the $x$-axis and the $Q$ quadrature the $y$-axis, with $I$ and $Q$ defined as in Eq.~(\ref{eqn-filt-I-Q}).

For the classification task, we choose the same pump strength $g_1=0.8g_1^{th}$, corresponding to operation at 80 \% of the instability threshold. This ensures that the two classes of quantum states exhibit identical squeezing strength. We set the phase of the pump tone $\phi_1^{(1)} = 0$ and $\phi_1^{(2)} = \pi$ to generate type-1 and type-2 states, respectively, such that the corresponding squeezed quadratures are orthogonal. A representative sampling of the measured outcomes in the $I$-$Q$ plane for these two parameter settings is shown in Fig.~\ref{fig:intro}(b), yielding quadrature angles $\pi/4$ and $-\pi/4$, as designed. The task is to distinguish these two squeezed states based only on the direction (or phase) of squeezing, and not the amount of squeezing. The simplicity of this example allows us to build intuition for how a quantum nonlinear amplifier performs the task and how its operating parameters can be optimized to maximize classification performance.

\section{\label{sec:analyzer-opt}Optimal system parameters for the classification task}

\subsection{\label{sec:perf-metrics}Performance metrics}

We first introduce two metrics to quantify classification performance based on the measured quadrature distributions (e.g., Fig.~\ref{fig:intro}(c)). The first performance metric is the mean separation between the distributions corresponding to the two types of states on the measured $I$-$Q$ quadrature plane $(\Delta\boldsymbol{\mu})$:
\begin{equation}
    \Delta\boldsymbol{\mu} = \begin{bmatrix}
        {\Delta I}
        \\
        {\Delta Q}
    \end{bmatrix}
    = \begin{bmatrix}
        \expect{I^{(1)}} - \expect{I^{(2)}}
        \\
        \expect{Q^{(1)}} - \expect{Q^{(2)}}
    \end{bmatrix},
\end{equation}
 where $I$ and $Q$ are defined in Eq.~(\ref{eqn-filt-I-Q}), and $||\Delta\boldsymbol{\mu}||$ is the $\ell^2$-norm of $\Delta\boldsymbol{\mu}$
\begin{equation}
    ||\Delta\boldsymbol{\mu}|| = \sqrt{\Delta I^2 + \Delta Q^2},
\end{equation}
For the classification of weak signals such as the ones we consider here, where the noise in the quadrature distributions is comparable to the mean separation, $||\Delta\mu||$ alone may not be an accurate proxy of empirical classification performance under finite sampling. In this case, which we will encounter for our particular task here, a more accurate metric of empirical classification performance is given by the Fisher discriminant, $D_F$ \cite{fisher_1936}. It is defined as follows:
\begin{equation}
    D_F = \Delta\boldsymbol{\mu}^T \cdot \mathbf{V}^{-1} \cdot \Delta\boldsymbol{\mu}.
\end{equation}
This metric also represents a natural multivariate generalization of the one-dimensional signal-to-noise ratio. The spread of the two distributions is captured by the combined variance of output fields of different types of states ($\mathbf{V}$),
\begin{equation}
    \mathbf{V} = \frac{1}{2}\left(\mathbf{\Sigma}_{I,Q}^{(1)} + \mathbf{\Sigma}_{I,Q}^{(2)} \right),
    \label{eq-V-mat}
\end{equation}
with
\begin{equation}
    \mathbf{\Sigma}_{I,Q}^{(i)} = \begin{bmatrix}
        \sigma_{I^{(i)}}^2 & \sigma_{I,Q^{(i)}}
        \\
        \sigma_{I,Q^{(i)}}
        & \sigma_{Q^{(i)}}^2
    \end{bmatrix},
\end{equation}
where
\begin{subequations}
    \begin{align}
        \sigma_{I^{(i)}}^2 &= \expect{I^{(i)^2}} - \expect{I^{(i)}}^2, \\
        \sigma_{I,Q^{(i)}}^2 &= \expect{I^{(i)}Q^{(i)}} - \expect{I^{(i)}}\expect{Q^{(i)}}, \\
        \sigma_{Q^{(i)}}^2 &= \expect{Q^{(i)^2}} - \expect{Q^{(i)}}^2.
    \end{align}
\end{subequations}

\begin{figure*}
    \centering
    \includegraphics[width=0.99\linewidth]{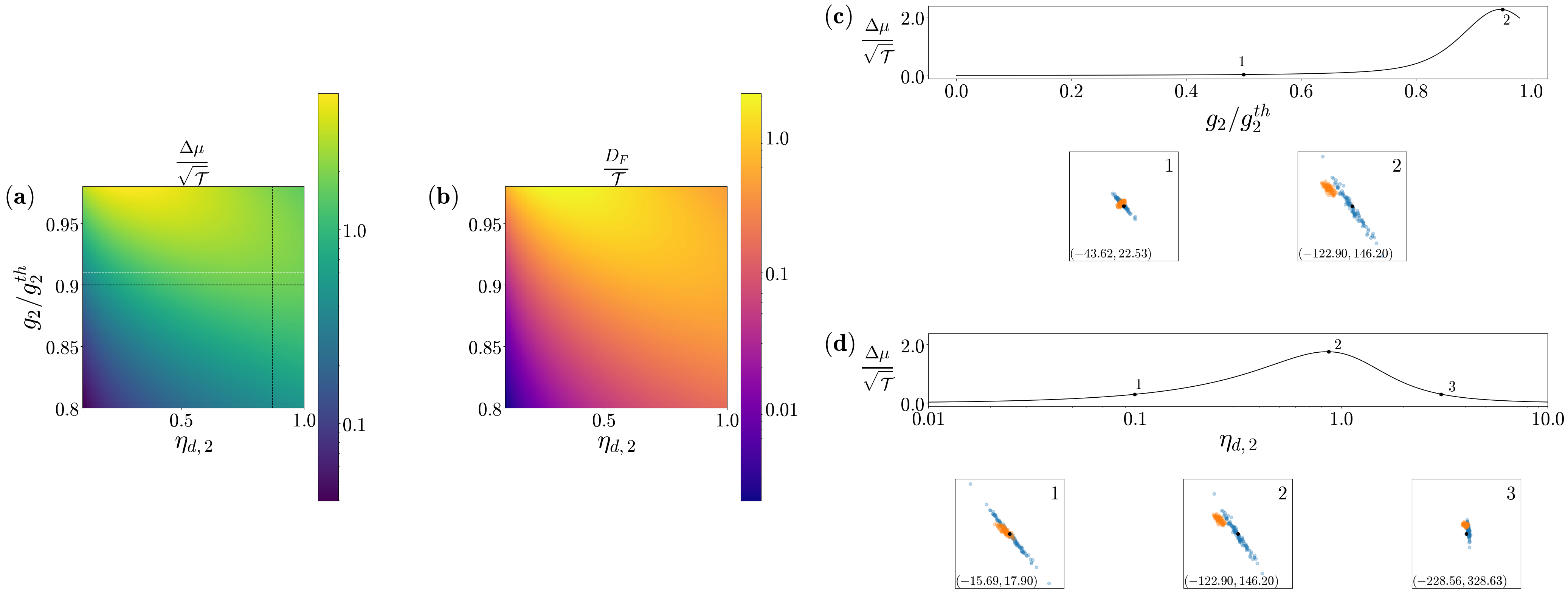}
    \caption{Variation of (a) normalized mean separation and (b) normalized Fisher discriminant versus pump strength $g_2$ (y-axis) and signal drive strength $\eta_{d,2}$ (x-axis). The vertical and horizontal black dashed lines in (a) indicate the cross sections shown in insets (c) and (d), respectively. The white dashed line marks the pump value giving 20 dB analyzer gain under the stiff-pump approximation. (c) Normalized mean separation versus $g_2$, along with measured quadratures for the two quantum-state types at two pump operating points shown in the insets below. The classification fidelity for instances 1 and 2 are 0.855 and 0.995 respectively. (d) Normalized mean separation versus $\eta_{d,2}$, with measured quadratures at three signal drive operating points shown in the insets below. The classification fidelity for instances 1, 2 and 3 are 0.845, 1 and 0.9 respectively. The coordinates at the bottom left corner of the instances in insets (c) and (d) correspond to the center of the plotted region of the $I$-$Q$ plot, all plots in each inset use identical axis ranges. The data shown is obtained by solving Eqs.~\ref{eq-app-teoms} and the data shown in the insets of (c) and (d) is obtained by solving Eqs.~\ref{eq-app-steoms} with 100 sample trajectories for each type of input state with integration time 800/$\kappa_2$.} 
    \label{fig:md-fd-strength}
\end{figure*}

\subsection{\label{sec:perf-metrics-comp}Method of computation}

To compute the mean and covariance of the output field for the different input states, we solve the truncated equations of motion (TEOMs) derived from the quantum master equation (QME), which corresponds to the infinite-shot limit of the stochastic quantum master equation (SQME)~\cite{wiseman_quantum_2009}. The means of the single-shot measurement outcomes are directly related to the quantum expectation values obtained from the TEOMs, as described below (see Appendix~\ref{app-filt} for the derivation):
\begin{equation}
    \expect{I} = \sqrt{\mathcal{T}} \avg{\hat{I}_2}, \qquad \expect{Q} = \sqrt{\mathcal{T}}\avg{\hat{Q}_2}.
\end{equation}
We estimate the measured quadrature covariance in terms of intracavity covariance of the output field, as if we have knowledge of the output states at the infinite shot limit,
\begin{equation}
    \mathbf{\Sigma}_{I,Q} \approx \begin{bmatrix}
          \Delta I_2& \Delta IQ_2 
        \\ \Delta IQ_2 
        & \Delta Q_2
        \label{eq-app-cov-mat}
    \end{bmatrix},
\end{equation}
with 
\begin{subequations}
    \begin{equation}
        \Delta I_2 = \avg{\hat{I}_2^2} - \avg{\hat{I}_2}^2,
    \end{equation}
    \begin{equation}
        \Delta IQ_2 = \avg{\hat{I}_2 \hat{Q}_2 + \hat{Q}_2 \hat{I}_2 }/2 - \avg{\hat{I}_2}\avg{\hat{Q}_2},
    \end{equation}
    \begin{equation}
        \Delta Q_2 = \avg{\hat{Q}_2^2} - \avg{\hat{Q}_2}^2.
    \end{equation}
\end{subequations}
This does not fully agree with the single-shot measurement covariance, but gives a good idea of where the performance optimum will occur, as they agree qualitatively. 

Before presenting the optimization results, we derive analytical expressions for the mean separation $\Delta\boldsymbol{\mu}$ and the combined variance matrix $\mathbf{V}$ that determine the Fisher discriminant $D_F $. 
\begin{equation}
    \Delta\boldsymbol{\mu} = i\sqrt{\Lambda \mathcal{T}} \mathbf{U} \bar{\mathbf{J}}^{-1} \begin{bmatrix} 2\Delta\bar{C}_{s_2^{\dagger}s_2}&  \Delta\bar{C}_{s_2s_2}  \\ - \Delta\bar{C}_{s_2^{\dagger}s_2^{\dagger}} & -2\Delta\bar{C}_{s_2^{\dagger}s_2} \end{bmatrix} \begin{bmatrix}\bar{s}_2 \\ \bar{s}^{\dagger}_2, \end{bmatrix},
    \label{eq-ms-pert}
\end{equation}
where  $\Delta\bar{C}_{o_1o_2} = \bar{C}_{o_1o_2}^{(2)} - \bar{C}_{o_1o_2}^{(1)}$ and with
\begin{equation}
     \bar{\mathbf{J}} = \begin{bmatrix}\left(i\Delta_2 - \frac{\kappa_2+\Gamma}{2} +i2 \bar{s}_2 \bar{s}_2^{\dagger}\right) & \left( -ig_2e^{i\phi_2} +i\bar{s}_2^{2} \right)  \\ \left( ig_2e^{-i\phi_2} -i\bar{s}_2^{\dagger^2} \right) & \left(-i\Delta_2 - \frac{\kappa_2+\Gamma}{2} -i2 \bar{s}_2 \bar{s}_2^{\dagger}\right)\end{bmatrix},
     \label{eq-ms-pert-J}
\end{equation}
and 
\begin{equation}
    \mathbf{U} = \frac{1}{\sqrt{2}}\begin{bmatrix}
    1 & 1 \\ -i &i
\end{bmatrix}.
\end{equation}
This result is obtained in the weak-nonlinearity limit using a nonstandard perturbative treatment of the TEOMs, with details provided in Appendix~\ref{app-pert}. The terms $\bar{s}_2$ and $\bar{C}_{s_2s_2},\bar{C}_{s_2^{\dagger}s_2} , \bar{C}_{s_2^{\dagger}s_2^{\dagger}}$ are zeroth-order contributions in the perturbative expansion for the mean and cumulant respectively. This analytical closed-form expression of the mean separation in a weak nonlinearity limit demonstrates some properties clearly. First, we see that it depends on $\sqrt{\Lambda}$ and $\sqrt{\mathcal{T}}$ linearly. That gives correctly in the limit that $\Lambda$ goes to zero, we have zero mean separation, thus classification fails. Secondly, the mean separation depends on the difference of the second order cumulants of the output distributions of different types. This shows how the nonlinear amplifier is able to compute the difference of cumulants via its quantum dynamics, and map them to first-order observables that are accessible via linear measurements. One can also calculate the Fisher discriminant under the perturbative approximation by making use of the approximate value of the combined variance matrix $\mathbf{V}$, which is obtained from Eq.~(\ref{eq-V-mat}) and Eq.~(\ref{eq-app-cov-mat}):
\begin{equation}
 \mathbf{V} = \frac{1}{4} \mathbf{U} \begin{bmatrix}\bar{C}_{s_2s_2}^{(1)} + \bar{C}_{s_2s_2}^{(2)} & \bar{C}_{s_2^{\dagger}s_2}^{(1)} + \bar{C}_{s_2^{\dagger}s_2}^{(2)} \\ \bar{C}_{s_2^{\dagger}s_2}^{(1)} + \bar{C}_{s_2^{\dagger}s_2}^{(2)} & \bar{C}_{s_2^{\dagger}s_2^{\dagger}}^{(1)} + \bar{C}_{s_2^{\dagger}s_2^{\dagger}}^{(2)} \end{bmatrix}\mathbf{U}^T.
 \label{eq-ms-pert-V}
\end{equation}
Lastly, in all the results we present, we analyze mean separation ($\Delta\mu/\sqrt{\mathcal{T}}$) and the Fisher discriminant ($D_F/\mathcal{T}$) normalized by filtering time. For steady-state operation, as the filtering time is increased, both of these performance metrics will increase. The scaling is clearly shown in Eq.~(\ref{eq-ms-pert}). This is also an intuitive result, as the noisy trajectories are integrated for a longer time, more information is collected, and one has more distinguishable output distributions.

\subsection{\label{sec:param-scan}Performance metrics under varying system parameters}

In this section, we focus on how the externally-adjustable system parameters affect the classification performance. These adjustable system parameters are identified as the pump strength ($g_2$) and phase ($\phi_2$), and the signal drive strength ($\eta_{d,2}$) and phase ($\phi_{d,2}$), where both drives are acting on the \textit{analyzer}. An arbitrary choice of parameters will not work, as the analyzer needs to be finely-tuned to a regime where nonlinear processing can be achieved. 

This section shows that optimal performance requires the joint optimization of the pump and signal drive strengths to identify the optimal operating points. Even with optimal drive strengths, an unfavorable choice of phase parameters can strongly degrade performance, making phase optimization equally essential. The remainder of this section provides an intuitive framework for selecting these optimal parameters by relating them directly to the classification task.

We separate the parameters into two groups, the strength parameters and the phase parameters, and show their effects separately. Let us start with the effect of the strength parameters on performance metrics. We show that the dependence of the mean separation, in \Fig{fig:md-fd-strength}(a), and the Fisher discriminant, in \Fig{fig:md-fd-strength}(b), are qualitatively very similar on these parameters. There is no global optimum in this two-dimensional parameter space. For an ideal linear amplifier, increasing the pump strength $g_2$ leads to improved performance, and the signal drive strength $\eta_{d,2}$ has no influence on performance. However, the Kerr nonlinearity of the practical \textit{analyzer} changes this notion. At any gain value, a certain drive strength is needed for the Kerr nonlinearity to have an effect; at larger gains, a smaller drive strength suffices as the amplification can increase the \textit{analyzer} mode occupation. However, it is also possible to have too strong a drive: then, the frequency shift due to the Kerr nonlinearity becomes too large and suppresses the useful nonlinear operation of the \textit{analyzer} mode by making it non-resonant. As a result, for every pump strength there is an optimum drive strength and vice versa, as can be seen clearly from the cross-sections of the 2D plots shown in Fig 2c, d.

\begin{figure}
    \centering
    \includegraphics[width=\linewidth]{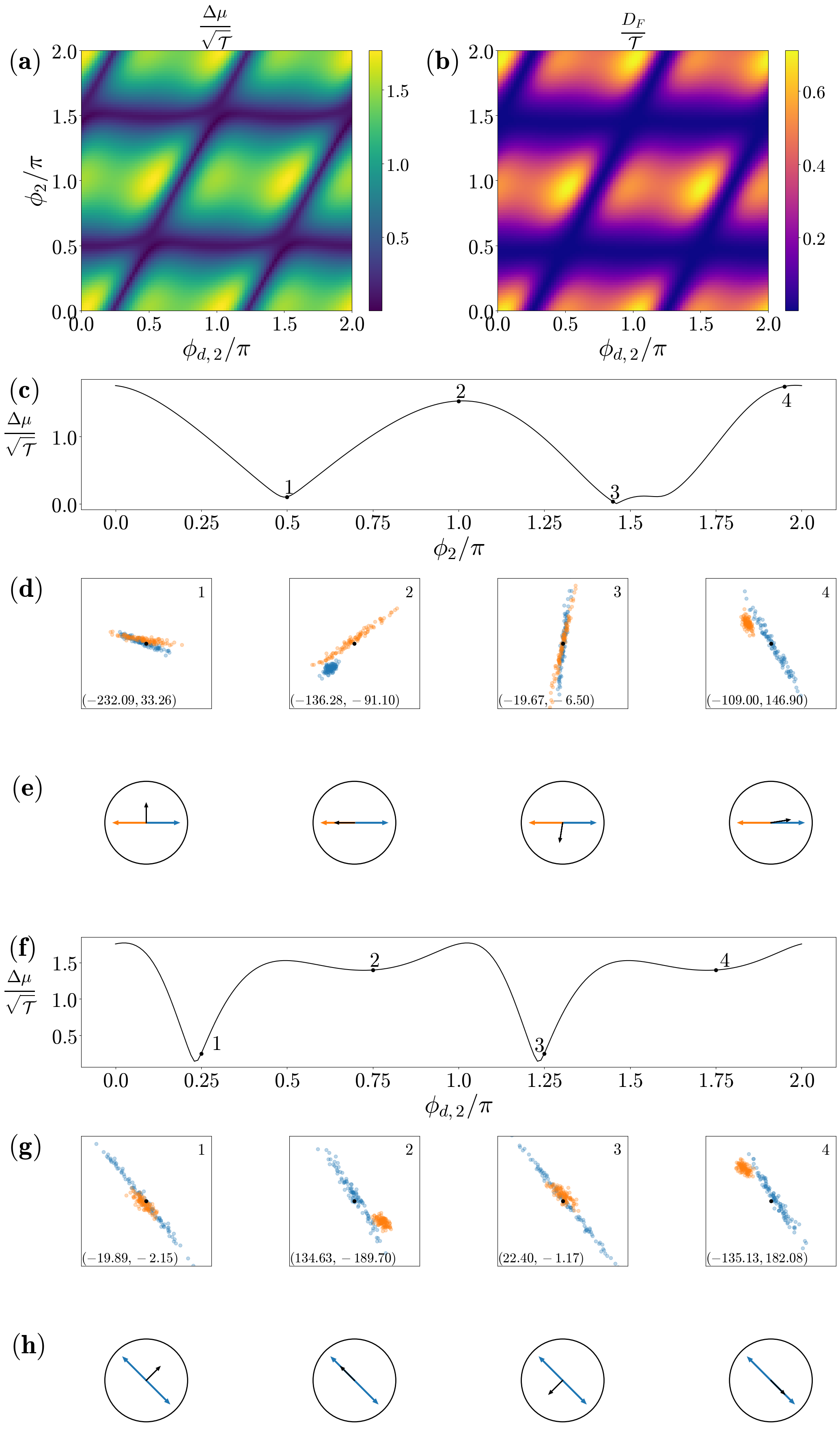}
    \caption{Variation of (a) normalized mean separation and (b) normalized Fisher discriminant versus pump phase $\phi_2$ (y-axis) and signal drive phase $\phi_{d,2}$ (x-axis). Inset (c) shows the normalized mean separation at fixed $\phi_{d,2}=0$ versus $\phi_2$, with inset (d) displaying corresponding measured quadratures for the two quantum state types at the operating points marked in (c). The classification fidelity for instances 1, 2, 3 and 4 are 0.81, 1, 0.785 and 1 respectively. Inset (e) illustrates the $\phi_1$ values on the unit circle generating type-1 and type-2 states (blue and orange), with the black arrow indicating $\phi_2$ at the chosen operating point. Inset (f) shows normalized mean separation at fixed $\phi_2=0$ versus $\phi_{d,2}$, and inset (g) shows the measured quadratures at the operating points marked in (f). The classification fidelity for instances 1, 2, 3 and 4 are 0.84, 0.99, 0.87 and 0.995 respectively. Inset (h) depicts the anti-squeezing (amplification) direction set by the fixed pump phase ($\phi_2 = 0$), aligned with the type-1 state (blue), with the black arrow marking the corresponding $\phi_{d,2}$. The coordinates at the bottom left corner of insets (d) and (g) correspond to the center of the plotted region of the $I$-$Q$ plot, all plots in each inset use identical axis ranges. The data shown is obtained by solving Eqs.~\ref{eq-app-teoms} and the data shown in the insets (d) and (g) is obtained by solving Eqs.~\ref{eq-app-steoms} with 100 sample trajectories for each type of input state with integration time 800/$\kappa_2$.}
    \label{fig:md-fd-phase}  
\end{figure}

We now focus on the phase parameters and their effect on the performance metrics. The dependence of the mean separation, shown in \Fig{fig:md-fd-phase}(a), and of the Fisher discriminant, shown in \Fig{fig:md-fd-phase}(b), is more complex than their respective dependence on strength parameters. First, there are multiple optimal points of operation in the parameter space of the two phases. Moreover, as in the case of strength parameters, the trends for the mean separation and the Fisher discriminant are similar: peak performance in terms of both metrics are achieved around similar parameter values. To analyze and understand the relation of these optimal phase parameters with the given task and the performance metrics, we look at the variation with respect to one of them at a time.

Looking at the mean separation against pump phase plot in \Fig{fig:md-fd-phase}(c), we see that the mean separation maxima, thus the best performance, are achieved when the phase of the \textit{analyzer} pump is the same as the phase of the \textit{squeezer} pump for either the type-1 (\Fig{fig:md-fd-phase}(e-4)) or type-2 (\Fig{fig:md-fd-phase}(e-2)) state ($\phi_2 = \phi_1^{(1)} = 0 \; or \; \phi_2 = \phi_1^{(2)} = \pi $), and the worst performance occurs when $\phi_2$ differs by $\pi/2$ from $\phi_1^{(1)}$ and $\phi_1^{(2)}$ [Fig.~\ref{fig:md-fd-phase}(e-1,3)]. This can be further quantified by the classification fidelity, which is unity at operating points 2 and 4, indicating no overlap between the two state distributions. In contrast, at operating points 1 and 3 the fidelity is 0.81 and 0.785, similar to the fidelity under linear operation, 0.84, as shown in \Fig{fig:intro}(c). The classification fidelity is evaluated using quadratic discriminant analysis on two distributions of 100 shots each at the corresponding operating point \cite{hastie_elements_of_statistical_09}.

We illustrate the mechanism using operating point 4 in \Fig{fig:md-fd-phase}(c), with output distributions shown in \Fig{fig:md-fd-phase}(d-4). Here, the analyzer pump phase matches the squeezer pump phase for the type-1 state and differs by $\pi$ for the type-2 state [\Fig{fig:md-fd-phase}(e-4)]. Consequently, the analyzer squeezes the type-1 state along the same quadrature, yielding a sharp elliptical output, while it acts as an anti-squeezer for the type-2 state, producing an almost circular distribution. This large covariance mismatch leads to a correspondingly large separation of the output means, since the mean separation is proportional to the difference in covariance of the two types of output states, as shown in Eq.~(\ref{eq-ms-pert}). Operating point 2 follows the same logic, where the analyzer pump phase aligns with the squeezer pump phase of the type-2 state instead of type-1. 

In contrast, at operating points 1 and 3 the analyzer phase is offset by $\pi/2$ from both squeezer phases [\Fig{fig:md-fd-phase}(e-1,3)], giving rise to similar output distributions for the two input states [\Fig{fig:md-fd-phase}(d-1,3)]. The resulting small covariance difference leads to a smaller mean separation.

Now, looking at the dependence of the mean separation on the signal drive phase in \Fig{fig:md-fd-phase}(f), we observe minima at $\phi_{d,2} = \pi/4$ and $\phi_{d,2} = 5\pi/4$ [\Fig{fig:md-fd-phase}(g-1,3)]. This occurs because this cross-section is taken at a pump phase of $\phi_2 = 0$, for which the anti-squeezing (amplification) axis lies along the $-\pi/4$ direction (and equivalently $3\pi/4$, as two ends of a line point in directions that are $\pi$ apart) in the $I$–$Q$ plane. At these signal drive phases, the direction of the signal drive is orthogonal to the amplification axis [\Fig{fig:md-fd-phase}(h-1,3)], resulting in a very small effective gain. Consequently, the input field is weakly amplified, leading to a vanishingly small mean separation, as the \textit{analyzer} mode population is insufficient for the Kerr nonlinearity to significantly affect the system dynamics. Consistent with this behavior, the classification fidelities at operating points 1 and 3 are 0.84 and 0.87, respectively, well below those at the optimal operating points 2 and 4 (0.99 and 0.995).

On the other hand, when the signal drive phase is aligned with the amplification direction (anti-squeezing direction) $\phi_{d,2} = 3\pi/4$ or $\phi_{d,2} = -\pi/4$ [\Fig{fig:md-fd-phase}(h-2,4)], we see a large value of the mean separation at those values and best performance is achieved around these points, with a twin-peak structure. The response $||\Delta\boldsymbol{\mu}||$ does not depend only on the linear response function of the analyzer, in which case it would exhibit only a single peak, but has a more complicated dependence on the covariance terms too, as indicated by Eq.~(\ref{eq-ms-pert}); this leads to a double-peak structure that has also been observed in previous work~\cite{khan_neural_2025}.

Lastly, one can see in the two-dimensional plots in \Fig{fig:md-fd-phase}(a) and \Fig{fig:md-fd-phase}(b), that all the optimal points maximizing both metrics occur around $\phi_2 = n \pi$. Around $\phi_2 = (2n+1) \pi/2$, both performance metrics take on their smallest values (close to zero), and there is not a big variation as the signal drive phase is varied. This shows that the optimization of phase parameters should start with identifying the optimal pump phase, followed by finding the optimal signal drive phase for that pump phase. In summary, therefore, the pump and signal drive strengths set the response of the \textit{analyzer} mode, with larger responses generally yielding improved classification performance. However, even for optimally-chosen amplitudes, a poor choice of phase parameters can drive both performance metrics close to zero, resulting in poor classification. Therefore a combined optimization of \textit{analyzer} pump phase and signal drive phase is crucial. 

\section{Effect of the classical measurement noise}
\label{sec:th}

To obtain a more robust readout with larger signal amplitude, the output of the quantum amplifier is further amplified by electronic amplifiers, such as HEMTs, later in the measurement chain. This electronic amplification introduces additional noise to the measured signal, which degrades the classification performance metrics. We parametrize the strength of this added noise by $\bar{n}_{\rm cl}$ in Eq.~(\ref{eqn-noisy-I-Qtraj}). The resulting covariance matrix of the measured quadratures under this noise is:
\begin{equation}
    \mathbf{\Sigma}_{I,Q}^{n} = \bar{n}_{\rm cl}\mathbb{I} + \mathbf{\Sigma}_{I,Q}.
\end{equation}
We refer to this contribution as \textit{classical noise} since it does not induce backaction on the quantum system and enters solely as additive noise on the measured quantum trajectories.
\begin{figure}
\includegraphics[width=\linewidth]{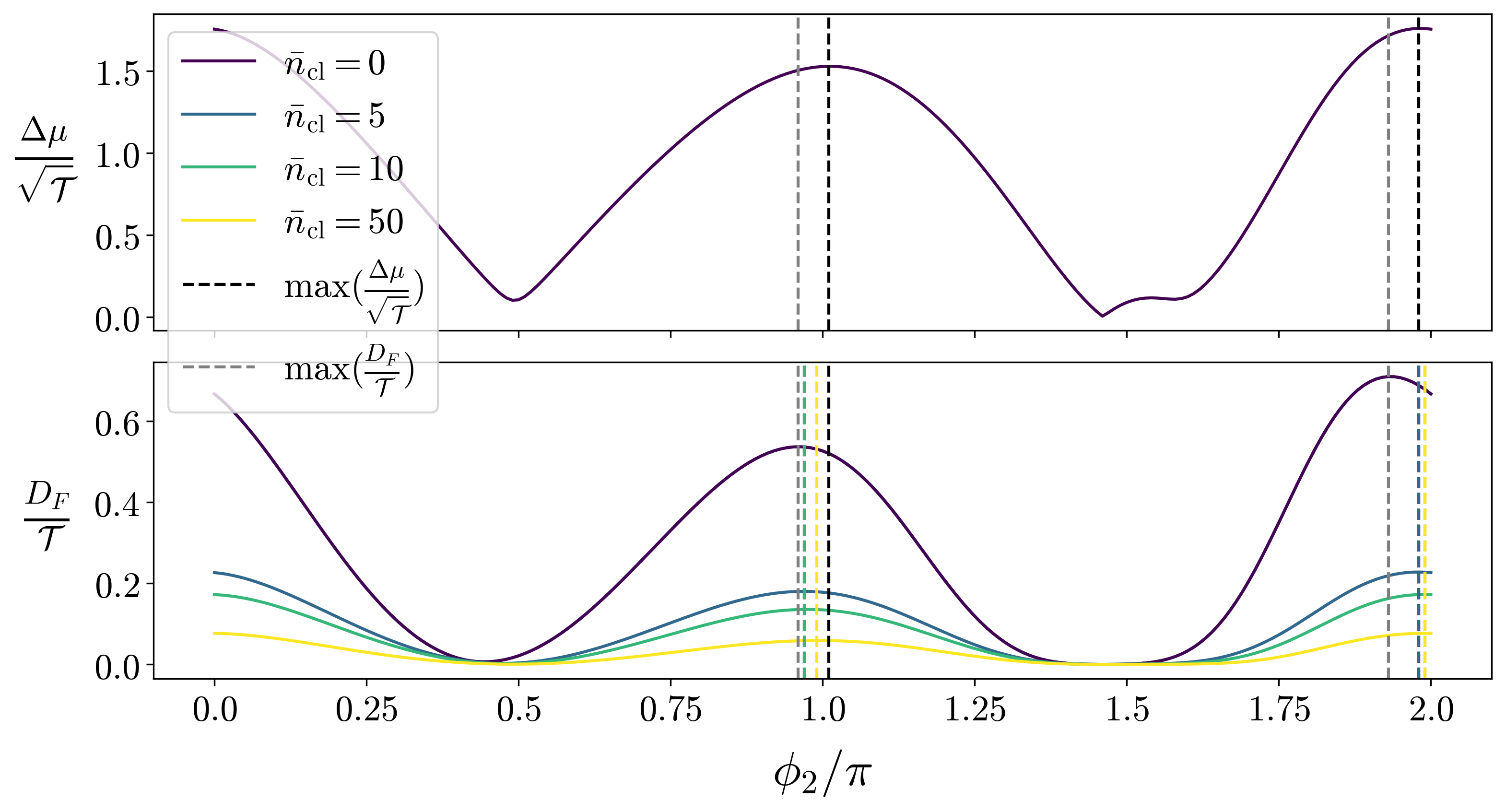}
\caption{\label{fig:md-fd-ncl} Normalized mean separation and the Fisher discriminant versus pump phase for different classical noise strengths (darker colors indicate larger noise). Black dashed lines mark the mean separation maxima, and grey dashed lines mark the Fisher discriminant maxima in the zero classical noise case. Colored dashed lines indicate the Fisher discriminant maxima for nonzero noise, using matching colors to the corresponding solid curves.}
\end{figure}
Using this covariance definition with added classical noise, we calculate the Fisher discriminant value for different noise contributions coming from the electronic amplification stage. We have plotted the variation of mean separation and the Fisher discriminant with different values of $\bar{n}_{cl}$  with respect to the pump phase in \Fig{fig:md-fd-ncl}. We note that the variation of the performance metrics with respect to parameters other than the pump phase is similar, and here we show results as a function of the pump phase for illustrative purposes.

The first observation one can make by looking at \Fig{fig:md-fd-ncl} is that the value of the Fisher discriminant decreases with increasing classical noise at all pump phase values. This is an expected result as this classical noise adds uncertainty equally along all directions in the $I$-$Q$ plane, thus reducing the signal-to-noise ratio (SNR) of the output independently of the system parameters.

We also see that the peaks of mean separation and the Fisher discriminant with zero classical noise, occur in similar neighborhoods of the pump phase but at different values. As the directionally uniform classical noise component in the system is increased, the pump phase value where the peak of the Fisher discriminant is achieved, shifts towards the one where mean separation peak is achieved. As this noise becomes the dominant noise in the system, the two different types of output distributions inevitably overlap with each other more, unless the mean separation between them is much greater than the variance of the added classical noise. In the presence of classical noise, therefore, to maximize the Fisher discriminant one should aim to maximize the mean separation.

\section{Application to qubit readout without cavity displacement}
\label{sec:qr}

\begin{figure}
    \centering
    \includegraphics[width=\linewidth]{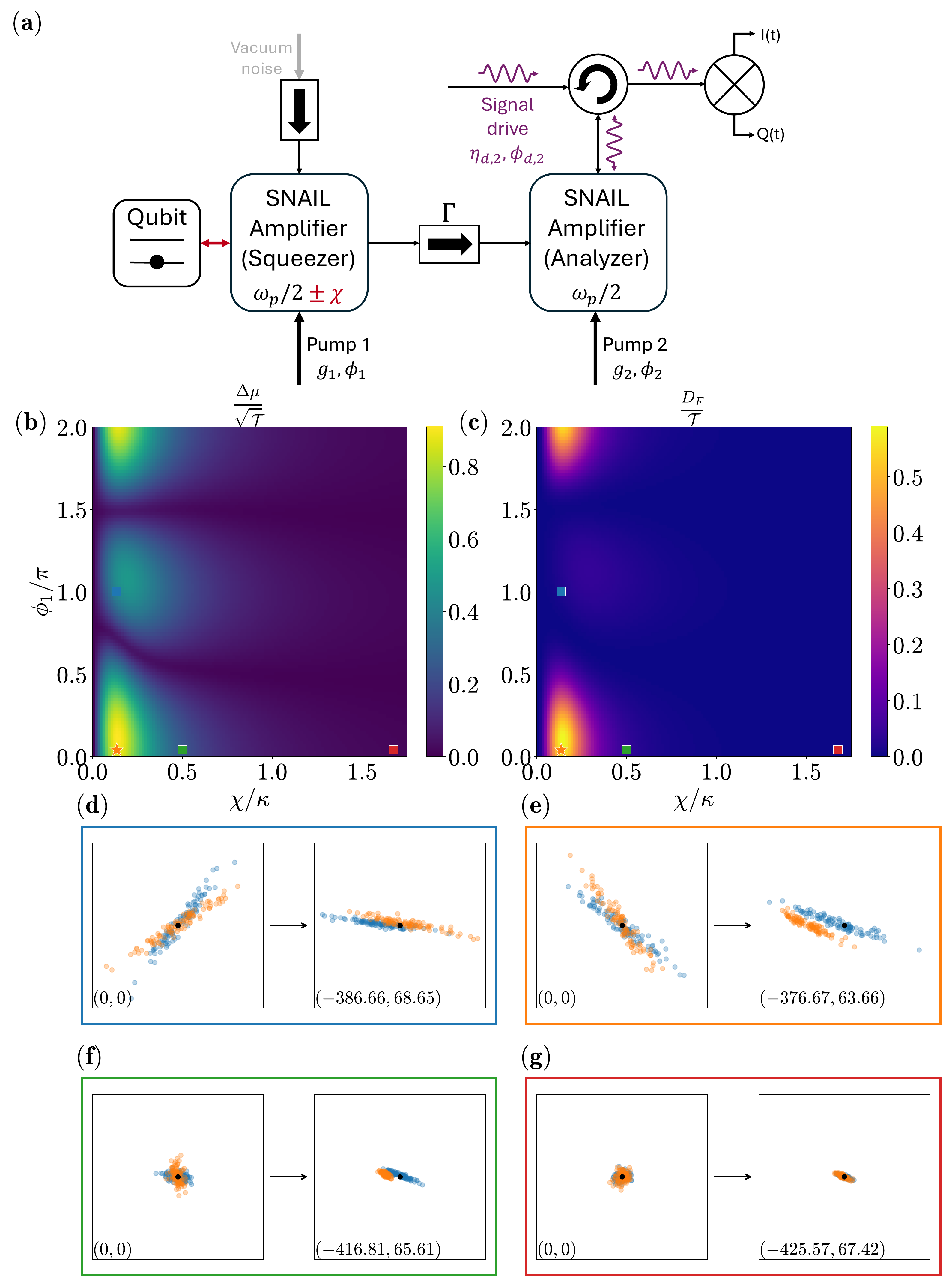}
    \caption{(a) Schematic of the modified \textit{squeezer}-\textit{analyzer} setup for qubit readout application without cavity displacement. The qubit is dispersively coupled to the \textit{squeezer} leading to a resonance shift $\chi$. Variation of (b) normalized mean separation and (c) normalized Fisher discriminant versus \textit{squeezer} pump phase ($\phi_1$) and qubit induced detuning ($\chi$). The input (on the left) and output (on the right) $I$-$Q$ plots as operating points marked in (b) and (c) are shown underneath with the matching colors in (d-g). The optimal operating point is shown with a star in orange and the corresponding $I$-$Q$ plots are shown in inset (e). The coordinates shown at the bottom-left corner of insets (d-g) correspond to the midpoint of the plotting window, all plots corresponding to input(output) states use identical axis ranges. The classification fidelity for output distributions in (d-g) are 0.905, 0.995, 0.995 and 0.61 respectively. The data shown in 
    (b) and (c) are obtained by using the analytical approximation in the perturbative limit in Eqs.~(\ref{eq-ms-pert}-\ref{eq-ms-pert-V}) and the data shown in the insets (d-g) is obtained by solving Eqs.~\ref{eq-app-steoms} with 100 sample trajectories for each type of input state with integration time 4000/$\kappa_2$. }
    \label{fig:qr}
\end{figure}

In this section, we discuss a possible application of our \textit{squeezer}-\textit{analyzer} architecture to qubit readout. We discuss a scenario where the information of the qubit state is imprinted onto squeezed vacuum states by dispersively coupling the qubit to the \textit{squeezer} as shown in Fig.~\ref{fig:qr}(a). The $analyzer$ performs state discrimination on these squeezed vacuum states via nonlinear amplification, effectively performing qubit readout. We consider this scenario to demonstrate how qubit readout can be performed by harnessing squeezing and nonlinearity as the main resources for quantum information processing, by considering simple modifications of the \textit{squeezer}-\textit{analyzer} system discussed in previous sections. 

For the dispersive qubit readout application without cavity displacement, we consider a simple setup in which the only change from our \textit{squeezer}-\textit{analyzer} setup is that the \textit{squeezer} is coupled to a qubit. In this setting, the detuning of the squeezer mode ($\Delta_1$) now takes on the value of the dispersive shift of the cavity resonance ($\chi$) imposed by the qubit state:
\begin{align}
\Delta_1 =
\begin{cases}
\begin{aligned}
 {}&\chi,  && \text{qubit in excited state ($\ket{1}$)} \\
 -{}&\chi, && \text{qubit in ground state ($\ket{0}$)}.
\end{aligned}
\end{cases}
\end{align}
With this change, the pointer states to be classified by the \textit{analyzer} will now be conditioned on the state of the qubit, instead of being set by an external pump phase, as we considered previously. As a result, classifying these pointer states will enable qubit state readout.

The advantage of this qubit readout scheme is that there is no displaced field in the resonator mode, which is coupled to the qubit; instead the resonator is squeezed by the \textit{squeezer} pump. Compared to standard dispersive qubit readout, this scheme offers potential advantages. In particular, operating with squeezed vacuum states can reduce the average intracavity photon number, therefore mitigating measurement-induced backaction \cite{blais_circuit_2021}. Beyond that, engineered squeezing may also positively influence qubit coherence as coupling a transmon to a squeezed field has been shown to enhance its transverse coherence time \cite{murch_reduction_2013}.

In this section, we analyze the optimal parameter regime for encoding the qubit state onto squeezed vacuum states. We analyze the effects of the dispersive shift ($\chi$) and the \textit{squeezer} pump phase ($\phi_1$) on optimal encoding that maximizes the classification performance when the analyzer is acting on the new pointer states generated by the qubit-\textit{squeezer} coupling.

In order to do so, we choose fixed pump strengths $g_1=0.9g_1^{th}$ and $g_2=0.9g_2^{th}$, we should note that the pump strength we have chosen for the \textit{squeezer} approximately corresponds to a photon number of 2 ($\langle\hat{s}_1^{\dagger}\hat{s}_1\rangle = 2.13$) verifying that the setup operates in a very low-photon regime. We also fix the \textit{analyzer} pump phase at $\phi_2=\pi/2$ and the analyzer signal drive phase at a value matching the amplification direction of the \textit{analyzer} amplifier ($\phi_{d,2} = -\pi/4 + \phi_2/2 = 0$). For a fixed signal drive strength on the analyzer ($\eta_{d,2} = 1$), we then sweep $\phi_1$ and $\chi$ and calculate the mean separation and Fisher discriminant semi-analytically using the perturbative expressions (Eqs.~\ref{eq-ms-pert}-\ref{eq-ms-pert-V}). We present the results in insets (b-c) of Fig.~\ref{fig:qr}. Moreover, we present the $I$-$Q$ plots of the input states (obtained by turning the analyzer off, i.e. setting $g_2=\eta_{d,2}=0$) and output states at various operating points in the insets (d-g) of of Fig.~\ref{fig:qr}.

Inspecting Fig.~\ref{fig:qr}(b-c), we notice that the maxima for both performance metrics are achieved in the same region of the parameter space around $\phi_1 \approx 0$ and $\chi \approx 0.2 \kappa$, where $\kappa = \kappa_1 + \Gamma$ is the total loss rate of the \textit{squeezer} mode including the bare resonator decay rate and the decay due to unidirectional coupler. 

First, the optimal value of $\phi_1$ depends on the fixed value of $\phi_2$, therefore it is better to address this as an optimal value for pump phase difference $\Delta\phi = \phi_2-\phi_1$. We have verified this by running the same simulation with different $\phi_2$ values while keeping the remaining fixed variables constant. In this case, the optimal value of $\phi_1$ shifts with $\phi_2$, with the optimal value of $\Delta\phi$ remaining the same. One can see how the performance deteriorates away from the optimal $\Delta\phi$, comparing the optimal $\Delta\phi$ point in inset (e) to  the non-optimal $\Delta\phi$ point in inset (d) in Fig.~\ref{fig:qr}.

Second, we observe that the optimal value of $\chi$ is smaller than the optimal dispersive shift for the standard dispersive qubit readout scheme at $\chi_{opt,disp} = \kappa/2$ \cite{blais_circuit_2021}(shown in inset Fig.~\ref{fig:qr}(f)) for phase-readout. The reason behind this is that in this scheme, as $\chi$ is increased, the \textit{squeezer} mode is tuned away from the single-mode-squeezed frequency, leading to a decrease in the squeezing strength. This can be seen from the input states in the Fig.~\ref{fig:qr}(e-g). At the same time as $\chi$ increases, the angle between the squeezing axes of the input states gets closer to being orthogonal. This feature of the encoding imposes an important trade-off between the degree of squeezing and the separation of the squeezing axes of the pointer states as $\chi$ is varied. 

For two single-mode Gaussian (squeezed) states with equal squeezing strength, their quantum distinguishability e.g. as quantified by the Helstrom bound or state overlap, is maximized when their squeezing ellipses are oriented along orthogonal quadrature axes \cite{curado_helstrom_2022}. We have analyzed such a binary discrimination task earlier in this paper. For the current qubit readout scheme where the qubit is dispersively coupled to the \textit{squeezer} mode, we can also find an operating regime where the two qubit-conditioned states of the squeezer have orthogonal squeezing axes. An example operating point in $(\phi_1,\chi)$ space where this can be achieved is marked in Fig.~\ref{fig:qr}(b) with a red marker corresponding to inset (g). However, for qubit-state encoding optimization, we find that the degree of squeezing and the separation of squeezing axes become competing factors that both affect classification accuracy. We see that if the states have orthogonal squeezing axes but the squeezing strength is not large enough, one ends up with poor classification as in Fig.~\ref{fig:qr}(g). We deduce that in this case, the orthogonal squeezing axes for the input states does not yield the optimal encoding, as the degree of squeezing to achieve the orthogonal separation becomes too small for effective state discrimination.

Lastly, the scheme we present here for encoding qubit states onto squeezed vacuum relies on a non-zero detuning of the squeezed mode, which fundamentally reduces the gain (equivalently the squeezing strength) in the \textit{squeezer}. Therefore, both normalized mean separation and normalized Fisher discriminant at the optimal performance points attain lower values for the qubit readout scheme without cavity displacement than for our earlier scheme. Therefore, the quantum trajectories need to be integrated for a longer time to achieve similar classification fidelity. Alternatively, one can increase the \textit{squeezer} pump strength ($g_1$), to increase performance metrics and therefore achieve similar fidelity with smaller integration time. However, we should note that increasing the squeezing strength also increases the average photon number in the readout cavity, therefore posing a trade-off.

\section{\label{sec:conc}Conclusion}

We have shown that quantum amplifiers operated beyond the linear regime can yield a measurable advantage for single-shot quantum state classification when the measurement chain is optimized for task-specific objectives rather than SNR alone. Our analysis is grounded in a realistic superconducting readout architecture comprising experimentally accessible components, including a SNAIL-based oscillator for state preparation and a second SNAIL oscillator functioning as a nonlinear cryogenic amplifier.

Within this framework, we identify optimal operating points for the externally tunable parameters of the nonlinear amplifier and provide a physical interpretation of these optima, clarifying the mechanisms that enhance classification fidelity. We further analyze the impact of downstream classical noise from an electronic amplifier and outline strategies to mitigate its effect, thereby preserving the performance gains enabled by nonlinear amplification.

Our results motivate deliberate operation of quantum amplifiers in the nonlinear regime and establish their potential as functional primitives for quantum information processing. While the present analysis focuses on performance metrics derived from second-order moments, it naturally extends to the processing of information encoded in higher-order moments of quantum signals, a regime of direct relevance to non-Gaussian quantum information protocols.

We further establish the feasibility of qubit readout without cavity displacement enabled by nonlinear amplification, showing that two zero-mean cavity states can be discriminated via phase information contained in their higher-order statistical moments, even when their first moments are identical. Building on this observation, we introduce a readout protocol that operates without cavity displacement and encodes the qubit state directly in these higher-order moments, thereby enabling discrimination in the zero-cavity-displacement regime. The present treatment does not account for measurement backaction, motivating future work that incorporates a realistic model of qubit–squeezer coupling. Moreover, the architecture we propose for qubit readout without cavity displacement is preliminary and could be further optimized, providing a natural direction for future work. More broadly, our findings provide a foundation for systematic, end-to-end, resource-constrained optimization of nonlinear quantum amplifiers for quantum information processing applications.

\begin{acknowledgments}

We thank Leon Bello for insightful discussions. We acknowledge support from the AFOSR under Grant No. FA9550-20-1-
0177 and the Army Research Office under Grant No. W911NF-23-1-0252. The views and conclusions contained in this document are those of the authors and
should not be interpreted as representing the official policies, either expressed or implied, of the Army Research
Office or the U.S. Government. The U.S. Government is
authorized to reproduce and distribute reprints for Government purposes, notwithstanding any copyright notation herein. Simulations in this paper were performed using the Princeton
Research Computing resources at Princeton University, which is a
consortium of groups led by the Princeton Institute for Computational Science and Engineering (PICSciE) and Office of Information Technology's Research Computing. B.M. acknowledges support from the NSF Graduate Research Fellowship Program Fellow No. 2022338601. \textbf{Conflict of Interest:} Michael Hatridge serves as a consultant for Quantum Circuits, Inc., receiving remuneration in the form of consulting fees, and hold equity in the form of stock options.
\end{acknowledgments}

\bibliography{squeezy-squeezy}
\newpage

\clearpage

\onecolumngrid
\appendix

\section{\label{app-full-derivation}Derivation of the system Hamiltonian in the most general form - Hamiltonian including third- and fourth-order nonlinearities of SNAIL}

Here, we want to establish the most general Hamiltonian to model a SNAIL amplifier including all effects of third- and fourth-order nonlinearities, driven at resonance frequency of the SNAIL ($\omega_s$) and twice of resonance frequency ($\omega_p = 2\omega_s$). We start with the Hamiltonian of the single SNAIL:
\begin{equation}
\begin{split}
    \hat{\mathcal{H}}_{SNAIL} = &\omega_s \hat{a}^{\dagger}\hat{a}  + g_3 \left(\hat{a}+\hat{a}^{\dagger}\right)^3 + g_4 \left(\hat{a}+\hat{a}^{\dagger}\right)^4 
    \\
    &+ \sqrt{\kappa_s}\epsilon_p \left(e^{-i\phi_p}e^{-i2\omega_s t} + e^{i\phi_p}e^{i2\omega_s t}\right) \left(\hat{a}+\hat{a}^{\dagger}\right) \\
    &+ \sqrt{\kappa_s}\eta_{sig} \left(e^{-i\phi_{sig}}e^{-i\omega_s t} + e^{i\phi_{sig}}e^{i\omega_s t}\right) \left(\hat{a}+\hat{a}^{\dagger}\right).
    \end{split}
\end{equation}
Then, we represent both the resonant SNAIL mode at $\omega_s$ and the pump mode at $2\omega_s$ within the mode $\hat{a}$.
\begin{align}
    \hat{a} = \hat{\tilde{s}} + \hat{\tilde{p}}.
\end{align}
We plug in this ansatz and go to a rotating frame with $\hat{s}=\hat{\tilde{s}}e^{-i\omega_st}$, $\hat{p}=\hat{\tilde{p}}e^{-i2\omega_st}$, where  $\omega_s = \omega_p/2$, and keep only non-rotating terms, and we end up with the following Hamiltonian:
\begin{equation}
    \begin{split}
        \hat{\mathcal{H}}_{3-4} = &\left(-\omega_s+24g_4\right)\hat{p}^{\dagger}\hat{p} + \left(24g_4\right) \hat{s}^{\dagger}\hat{s} + 3g_3\left(\hat{p}^{\dagger}\hat{s}\hat{s} + \hat{p}\hat{s}^{\dagger}\hat{s}^{\dagger} \right) + 6g_4\left(\hat{p}^{\dagger}\hat{p}^{\dagger}\hat{p}\hat{p} + \hat{s}^{\dagger}\hat{s}^{\dagger}\hat{s}\hat{s}\right) + 24g_4\left(\hat{p}^{\dagger}\hat{p}\hat{s}^{\dagger}\hat{s}\right)
        \\
        &+\sqrt{\kappa_s}\epsilon_p\left(e^{-i\phi_p}\hat{p}^{\dagger} + e^{i\phi_p}\hat{p}\right) +\sqrt{\kappa_s}\eta_{sig}\left(e^{-i\phi_{sig}}\hat{s}^{\dagger} + e^{i\phi_{sig}}\hat{s}\right).
    \end{split}
\end{equation}
If we assume the pump is stiff, the Hamiltonian will be simplified in the following way:
\begin{equation}
    \begin{split}
        \hat{\mathcal{H}}_{3-4}^{stiff} =  &\left(24g_4\left(1+|\Bar{\mathcal{P}}|^2\right)\right) \hat{s}^{\dagger}\hat{s} + 3g_3\left(\Bar{\mathcal{P}}^* \hat{s}\hat{s} + \Bar{\mathcal{P}}\hat{s}^{\dagger}\hat{s}^{\dagger} \right) 
        + 6g_4 \hat{s}^{\dagger}\hat{s}^{\dagger}\hat{s}\hat{s} +\sqrt{\kappa_s}\eta_{sig}\left(e^{-i\phi_{sig}}\hat{s}^{\dagger} + e^{i\phi_{sig}}\hat{s}\right).
    \end{split}
\end{equation}
with $\Bar{\mathcal{P}} = \frac{i\sqrt{\kappa_s}\epsilon_pe^{-i\phi_p}}{\left(i\omega_s - \frac{\kappa_s}{2}\right)}$.

When we match the parameters here with the \textit{analyzer} in the main text we have:
\begin{subequations}
\label{eqn-param-conv}
 \begin{equation}
     \Delta_2 = -24g_4\left(1+|\bar{\mathcal{P}}|^2\right),
     \label{eqn-detuning-conv}
 \end{equation}
 \begin{equation}
     \Lambda = -12 g_4,
     \label{eqn-lambda-conv}
 \end{equation}
 \begin{equation}
     \epsilon_{p} = \left|-\frac{g_2e^{\phi_2}}{i6g_{3}\sqrt{\kappa_2}\chi_{2}}\right|,
     \label{eqn-pump-strength-conv}
 \end{equation}
 \begin{equation}
     \phi_{p} = -{\rm arg}\left\{ -\frac{g_2e^{\phi_2}}{i6g_{3}\sqrt{\kappa_j}\chi_{2}}\right\},
     \label{eqn-pump-phase-conv}
 \end{equation}
 \begin{equation}
     \eta_{d,2} = \eta_{sig},
     \label{eqn-drive-strength-conv}
 \end{equation}
 \begin{equation}
     \phi_{sig} = \phi_{d,2} - \pi/2,
     \label{eqn-drive-phase-conv}
 \end{equation}
\end{subequations}
with $\chi_2^{-1} = -i\Delta_2+\frac{\kappa_2}{2}$.

We can also model the system in the main text, consisting of an \textit{squeezer} SNAIL unidirectionally coupled to an \textit{analyzer} SNAIL, in this way. The Liouvillian can be split into three parts as follows:
\begin{align}
    \mathcal{L}_{non-stiff} = \mathcal{L}_{sq-non-stiff} + \mathcal{L}_{an-non-stiff} + \mathcal{L}_{coup},
\end{align}
and $\mathcal{L}_{coup}$ is as in Eq.~(\ref{eqn-coup-L}). The \textit{squeezer} Liouvillian is modified to include non-stiff dynamics to:
\begin{align}
    \mathcal{L}_{sq}^{full}\hat{\rho} = -i[\hat{\mathcal{H}}_{sq}^{full},\hat{\rho}] + \kappa_1\mathcal{D}[\hat{s}_1+\hat{p}_1]\hat{\rho},
\end{align}
with
\begin{equation}
    \begin{split}
    \hat{\mathcal{H}}_{sq}^{full} = -\omega_{s,1} \hat{p}_1^{\dagger}\hat{p}_1 + 3g_{3,1}\left(\hat{p}_1^{\dagger}\hat{s}_1\hat{s}_1 + \hat{p}_1 \hat{s}_1^{\dagger} \hat{s}_1^{\dagger} \right) +\sqrt{\kappa_1}\epsilon_{p,1}\left(e^{-i\phi_{p,1}}\hat{p}_1^{\dagger} + e^{i\phi_{p,1}}\hat{p}_1\right).
    \end{split}
\end{equation}
Finally, the \textit{analyzer} Liouvillian is as follows:
\begin{align}
    \mathcal{L}_{an}^{full}\hat{\rho} = -i[\hat{\mathcal{H}}_{an}^{full},\hat{\rho}] + \kappa_2\mathcal{D}[\hat{s}_2+\hat{p}_2]\hat{\rho},
\end{align}
with
\begin{equation}
    \begin{split}
        \hat{\mathcal{H}}_{an}^{full} = &\left(-\omega_{s,2}+24g_4\right)\hat{p}_2^{\dagger}\hat{p}_2 + \left(24g_4\right) \hat{s}_2^{\dagger}\hat{s}_2 + 3g_{3,2}\left(\hat{p}_2^{\dagger}\hat{s}_2 \hat{s}_2 + \hat{p}_2 \hat{s}_2^{\dagger} \hat{s}_2^{\dagger} \right) + 6g_4\left(\hat{p}_2^{\dagger} \hat{p}_2^{\dagger} \hat{p}_2 \hat{p}_2 + \hat{s}_2^{\dagger} \hat{s}_2^{\dagger} \hat{s}_2 \hat{s}_2\right) 
        \\
        &+ 24g_4\left(\hat{p}_2^{\dagger} \hat{p}_2 \hat{s}_2^{\dagger} \hat{s}_2 \right) +\sqrt{\kappa_2}\epsilon_{p,2}\left(e^{-i\phi_{p,2}}\hat{p}_2^{\dagger} + e^{i\phi_{p,2}}\hat{p}_2\right) +\sqrt{\kappa_2}\eta_{d,2}\left(e^{-i\phi_{sig,2}} \hat{s}_2^{\dagger} + e^{i\phi_{sig,2}} \hat{s}_2 \right).
    \end{split}
\end{equation}
One can similarly match the pump strength and phase for the \textit{squeezer} with the stiff-pump approximation as in Eq.~(\ref{eqn-pump-strength-conv}) and Eq.~(\ref{eqn-pump-phase-conv}), simply by changing indices 2 to 1. 

\section{\label{app-cumulants}Cumulant expansion method and STEOMs of the system}

In this section, we explain the relation between the moments of a multimode quantum system and their cumulants and then show how we use this relation to derive a finite set of equations of motions of the cumulants of our system. An n-order moment of an arbitrary multimode system can be written in terms of cumulants as follows:
\begin{equation}
    \avg{o_1 o_2 \cdots o_n} = \sum_{\pi} \prod_{{B} \in \pi} C_{{o_i}: i \in {B}},
\end{equation}
where $\pi$ defines the set of possible partitions of operators in the n-order moment, and $B$ indicates elements in this
set of partitions.

We show this on an example derived from the SQME describing our system, which is introduced in the main text in Eq.~(\ref{eq-s2-STEOM}). If we derive the equation of motion for $\avg{\hat{s}_2}^c = tr\{\mathcal{L}\hat{\rho}^c\}$ from the SQME we end up with the following equation:
\begin{equation}
    \begin{split}
    d\avg{\hat{s}_2}^c = & \left( -{\Gamma_1}\avg{\hat{s}_1}^c + \left(i\Delta_2 -\frac{\kappa_2+\Gamma_1}{2}\right) \avg{\hat{s}_2}^c +i\Lambda \avg{\hat{s}_2^{\dagger}\hat{s}_2\hat{s}_2}^c  -ig_2e^{i\phi_2} \avg{\hat{s}_2^{\dagger}}^c -\sqrt{\kappa_2}\eta_{d,2}e^{i\phi_{d,2}}\right) dt 
    \\
    &+ \sqrt{\frac{\kappa_2}{2}}\left(   \avg{\hat{s}_2\hat{s}_2}^c + \avg{\hat{s}_2^{\dagger}\hat{s}_2}^c - \avg{\hat{s}_2}^c\avg{\hat{s}_2}^c - \avg{\hat{s}_2^{\dagger}}^c\avg{\hat{s}_2}^c \right)dW_I \\
    &-i \sqrt{\frac{\kappa_2}{2}}\left(   \avg{\hat{s}_2\hat{s}_2}^c - \avg{\hat{s}_2^{\dagger}\hat{s}_2}^c - \avg{\hat{s}_2}^c\avg{\hat{s}_2}^c + \avg{\hat{s}_2^{\dagger}}^c\avg{\hat{s}_2}^c \right)dW_Q.
    \end{split}
\end{equation}
We notice two things from here. First, the measurement operator couples this first-order moment equation to second-order moments. We know $C_{o_1o_2} = \avg{o_1 o_2} - \avg{o_1}\avg{o_2}$, so we can replace the terms of second-order moments with cumulants in the measurement terms. 

The second and more important observation is that the Kerr term (associated with $\Lambda$), couples this first-order moment equation to a third-order moment. This coupling to the higher-order moments will lead to an unclosed set of equations as the equations of motions lower-order moments include terms with higher-order moments. In order to prevent this and end up with finite number of equations, we truncate our cumulants at second-order, meaning we assume cumulants of order three and higher are all zero. This approximation means that we assume that our states are Gaussian, which is a good approximation in the case of weakly nonlinear systems with Gaussian inputs, as in our case. We now show how this is done using the third-order moment associated with the Kerr term. First, we write it in terms of cumulants:
\begin{equation}
    \avg{\hat{s}_2^{\dagger}\hat{s}_2\hat{s}_2}^c = C_{\hat{s}_2^{\dagger}\hat{s}_2\hat{s}_2}^c + 2C_{\hat{s}_2^{\dagger}\hat{s}_2}^c\avg{\hat{s}_2}^c + C_{\hat{s}_2\hat{s}_2}^c\avg{\hat{s}_2^{\dagger}}^c +\avg{\hat{s}_2^{\dagger}}^c\avg{\hat{s}_2}^{c}\avg{\hat{s}_2}^{c} .
\end{equation}
Then we assume that the third-order moment is zero ($C_{\hat{s}_2^{\dagger}\hat{s}_2\hat{s}_2}^c = 0$), thus we can write the third-order cumulant in terms of second and first-order cumulants. After doing these conversions we end up with the following equation which only contains first and second-order cumulants:
\begin{equation}
    \begin{split}
    d\avg{\hat{s}_2}^c = & dt \Bigg( -{\Gamma_1}\avg{\hat{s}_1}^c + \left(i\Delta_2 -\frac{\kappa_2+\Gamma_1}{2}\right) \avg{\hat{s}_2}^c +i\Lambda\left(2 C^c_{\hat{s}_2^{\dagger}\hat{s}_2} \avg{\hat{s}_2}^c + \avg{\hat{s}_2^{\dagger}}^c\left( \avg{\hat{s}_2}^{c^2} +  C^c_{\hat{s}_2\hat{s}_2}\right) \right) -ig_2e^{i\phi_2} \avg{\hat{s}_2^{\dagger}}^c
    \\
     & - \sqrt{\kappa_2}\eta_{d,2}e^{i\phi_{d,2}}\Bigg) 
    + \sqrt{\frac{\kappa_2}{2}}\left(   C^c_{\hat{s}_2\hat{s}_2} +  C^c_{\hat{s}_2^{\dagger}\hat{s}_2} \right)dW_I -i \sqrt{\frac{\kappa_2}{2}}\left(   C^c_{\hat{s}_2\hat{s}_2} -  C^c_{\hat{s}_2^{\dagger}\hat{s}_2} \right)dW_Q.
    \end{split}
\end{equation}
We apply the same procedure to all the other first- and second-order moments' equations of motion and end up with 15 coupled closed set of equations, STEOMs, for this system:
\begin{subequations}
    \begin{equation}
    d\avg{\hat{s}_1}^c = dt\Bigg(-\frac{\Gamma_1+\kappa_1}{2} \avg{\hat{s}_1}^c +i\Delta_1 \avg{\hat{s}_1}^c -ig_1e^{i\phi_1} \avg{\hat{s}_1^{\dagger}}^c \Bigg) + \sqrt{\frac{\kappa_2}{2}}\left( C^c_{s_1s_2} +  C^c_{s_2^{\dagger}s_1}\right)dW_I + -i \sqrt{\frac{\kappa_2}{2}}\left( C^c_{s_1s_2} -  C^c_{s_2^{\dagger}s_1}\right)dW_Q,
\end{equation}
\begin{equation}
    \begin{split}
    d\avg{\hat{s}_2}^c = & dt\Bigg( -{\Gamma_1}\avg{\hat{s}_1}^c + \left(i\Delta_2 -\frac{\kappa_2+\Gamma_1}{2}\right) \avg{\hat{s}_2}^c +i\Lambda\left(2 C^c_{s_2^{\dagger}s_2} \avg{\hat{s}_2}^c + \avg{\hat{s}_2^{\dagger}}^c\left( \avg{\hat{s}_2}^{c^2} +  C^c_{s_2s_2}\right) \right) -ig_2e^{i\phi_2} \avg{\hat{s}_2^{\dagger}}^c
    \\
    &-\sqrt{\kappa_2}\eta_{d,2}e^{i\phi_{d,2}}\Bigg) 
    + \sqrt{\frac{\kappa_2}{2}}\left(   C^c_{s_2s_2} +  C^c_{s_2^{\dagger}s_2} \right)dW_I -i \sqrt{\frac{\kappa_2}{2}}\left(   C^c_{s_2s_2} -  C^c_{s_2^{\dagger}s_2} \right)dW_Q,
    \end{split}
\end{equation}
\begin{equation}
d\avg{\hat{s}_1^{\dagger}}^c = \Bigg(-\frac{\Gamma_1+\kappa_1}{2} \avg{\hat{s}_1^{\dagger}}^c -i\Delta_1 \avg{\hat{s}_1^{\dagger}}^c +ig_1e^{-i\phi_1} \avg{\hat{s}_1}^c \Bigg) dt + \sqrt{\frac{\kappa_2}{2}}\left( C^c_{s_1^{\dagger}s_2} +  C^c_{s_1^{\dagger}s_2^{\dagger}}\right)dW_I -i \sqrt{\frac{\kappa_2}{2}}\left( C^c_{s_1^{\dagger}s_2} -  C^c_{s_1^{\dagger}s_2^{\dagger}}\right)dW_Q,
\end{equation}
\begin{equation}
    \begin{split}
    d\avg{\hat{s}_2^{\dagger}}^c = &dt\Bigg( -{\Gamma_1}\avg{\hat{s}_1^{\dagger}}^c -\left(i\Delta_2 +\frac{\kappa_2+\Gamma_1}{2} \right) \avg{\hat{s}_2^{\dagger}}^c -i\Lambda\left(2 C^c_{s_2^{\dagger}s_2} \avg{\hat{s}_2^{\dagger}}^c + \avg{\hat{s}_2}^c\left(\avg{\hat{s}_2^{\dagger}}^{c^2} +C^c_{s_2^{\dagger}s_2^{\dagger}}\right) \right) +ig_2e^{-i\phi_2} \avg{\hat{s}_2}^c  
    \\
     &-\sqrt{\kappa_2}\eta_{d,2}e^{-i\phi_{d,2}}\Bigg)
    + \sqrt{\frac{\kappa_2}{2}}\left(   C^c_{s_2^{\dagger}s_2} +  C^c_{s_2^{\dagger}s_2^{\dagger}} \right)dW_I -i \sqrt{\frac{\kappa_2}{2}}\left(   C^c_{s_2^{\dagger}s_2} -  C^c_{s_2^{\dagger}s_2^{\dagger}} \right)dW_Q,
    \end{split}
\end{equation}
\begin{equation}
    \begin{split}
        \dot{ C^c_{s_1s_1}} &= (-\Gamma_1-\kappa_1+i2\Delta_1) C^c_{s_1s_1} -2 \kappa_2  C^c_{s_1s_2}   C^c_{s_2^{\dagger}s_1}  - ig_1e^{i\phi_1}(1+2 C^c_{s_1^{\dagger}s_1}),
    \end{split}
\end{equation}
\begin{equation}
    \begin{split}
        \dot{ C^c_{s_1s_2}} = &-\Gamma_1 C^c_{s_1s_1} + \left(-\Gamma_1 - \frac{\kappa_1+\kappa_2}{2} +i(\Delta_1 + \Delta_2)\right)C^c_{s_1s_2}  -\kappa_2 C^c_{s_1s_2}C^c_{s_2^{\dagger}s_2} -\kappa_2 C^c_{s_2s_2}C^c_{s_2^{\dagger}s_1} 
        \\
        &+i\Lambda\left(C^c_{s_2^{\dagger}s_1}C^c_{s_2s_2}+ 2C^c_{s_1s_2}C^c_{s_2^{\dagger}s_2} + C^c_{s_2^{\dagger}s_1}\avg{\hat{s}_2}^{c^2} + 2C^c_{s_1s_2}\avg{\hat{s}_2}^c\avg{\hat{s}_2^{\dagger}}^c\right) 
        -ig_1e^{i\phi_1}C^c_{s_1^{\dagger}s_2} -ig_2e^{i\phi_2}C^c_{s_2^{\dagger}s_1},
    \end{split}
\end{equation}
\begin{equation}
    \begin{split}
        \dot{ C^c_{s_2s_2}} = &-2\Gamma_1 C^c_{s_1s_2} + (-\Gamma_1-\kappa_2 +i2\Delta_2)C^c_{s_2s_2} -2\kappa_2 C^c_{s_2^{\dagger}s_2}C^c_{s_2s_2} 
        \\
        &+ i\Lambda\left(C^c_{s_2s_2} + 6C^c_{s_2s_2}C^c_{s_2^{\dagger}s_2} + \avg{\hat{s}_2}^{c^2} + 2C^c_{s_2^{\dagger}s_2} \avg{\hat{s}_2}^{c^2} + 4C^c_{s_2s_2} \avg{\hat{s}_2}^c \avg{\hat{s}_2^{\dagger}}^c \right) -ig_2e^{i\phi_2}\left(1+2C^c_{s_2^{\dagger}s_2}\right),  
    \end{split}
\end{equation}
\begin{equation}
    \begin{split}
        \dot{ C^c_{s_1^{\dagger}s_1}} = (-\Gamma_1-\kappa_1)C^c_{s_1^{\dagger}s_1} - \kappa_2 C^c_{s_1^{\dagger}s_2}C^c_{s_2^{\dagger}s_1} - \kappa_2 C^c_{s_1s_2}C^c_{s_1^{\dagger}s_2^{\dagger}} -i g_1 e^{i\phi_1} C^c_{s_1^{\dagger}s_1^{\dagger}} + ig_1 e^{-i\phi_1} C^c_{s_1 s_1},
    \end{split}
\end{equation}
\begin{equation}
    \begin{split}
        \dot{ C^c_{s_1^{\dagger}s_2}} = & -\Gamma_1 C^c_{s_1^{\dagger}s_1} + (-\Gamma_1 - \frac{\kappa_1+\kappa_2}{2} + i(-\Delta_1+\Delta_2))C^c_{s_1^{\dagger}s_2} - \kappa_2 C^c_{s_1^{\dagger}s_2}C^c_{s_2^{\dagger}s_2} -\kappa_2 C^c_{s_1^{\dagger}s_2^{\dagger}} C^c_{s_2 s_2} 
        \\
        &+ i\Lambda \left(2C^c_{s_1^{\dagger}s_2}C^c_{s_2^{\dagger}s_2} + C^c_{s_1^{\dagger}s_2^{\dagger}}C^c_{s_2 s_2} + C^c_{s_1^{\dagger}s_2^{\dagger}}\avg{\hat{s}_2}^{c^2} + 2 C^c_{s_1^{\dagger}s_2}\avg{\hat{s}_2}^c\avg{\hat{s}_2^{\dagger}}^c\right) + ig_1e^{-i\phi_1}C^c_{s_1s_2} -ig_2e^{i\phi_2} C^c_{s_1^{\dagger}s_2^{\dagger}},
    \end{split}
\end{equation}
\begin{equation}
    \begin{split}
        \dot{ C^c_{s_2^{\dagger}s_1}} =  &-\Gamma_1 C^c_{s_1^{\dagger}s_1} + \left(-\Gamma_1 -\frac{\kappa_1+\kappa_2}{2} + i(\Delta_1 - \Delta_2) \right) C^c_{s_2^{\dagger}s_1} - \kappa_2C^c_{s_2^{\dagger}s_1}C^c_{s_2^{\dagger}s_2} - \kappa_2 C^c_{s_1s_2}C^c_{s_2^{\dagger}s_2^{\dagger}}
        \\
        &-i\Lambda\left( 2C^c_{s_2^{\dagger}s_1}C^c_{s_2^{\dagger}s_2} +C^c_{s_1s_2}C^c_{s_2^{\dagger}s_2^{\dagger}}+ 2C^c_{s_2^{\dagger}s_1}\avg{\hat{s}_2}^c\avg{\hat{s}_2^{\dagger}}^c + C^c_{s_1s_2} \avg{\hat{s}_2^{\dagger}}^{c^2} \right) + ig_2e^{-i\phi_2} C^c_{s_1s_2} -ig_1e^{i\phi_1}C^c_{s_1^{\dagger}s_2^{\dagger}},
    \end{split}
\end{equation}
\begin{equation}
    \begin{split}
        \dot{ C^c_{s_2^{\dagger}s_2}} = &-\Gamma_1 C^c_{s_1^{\dagger}s_2} -\Gamma_1 C^c_{s_2^{\dagger}s_1} + (-\Gamma_1-\kappa_2)C^c_{s_2^{\dagger}s_2} - \kappa_2 C^c_{s_2^{\dagger}s_2} C^c_{s_2^{\dagger}s_2} -\kappa_2 C^c_{s_2s_2}C^c_{s_2^{\dagger}s_2^{\dagger}} 
        \\
        &+ i\Lambda\left(C^c_{s_2^{\dagger}s_2^{\dagger}} \avg{\hat{s}_2}^{c^2} - C^c_{s_2s_2}\avg{\hat{s}_2^{\dagger}}^{c^2}\right) -ig_2e^{i\phi_2} C^c_{s_2^{\dagger}s_2^{\dagger}}+ig_2e^{-i\phi_2} C^c_{s_2s_2},
    \end{split}
\end{equation}
\begin{equation}
    \begin{split}
        \dot{ C^c_{s_1^{\dagger}s_1^{\dagger}}} = (-\Gamma_1-\kappa_1-i2\Delta_1) C^c_{s_1^{\dagger}s_1^{\dagger}} - 2\kappa_2 C^c_{s_1^{\dagger}s_2}C^c_{s_1^{\dagger}s_2^{\dagger}} + ig_1e^{-i\phi_1}(1+2C^c_{s_1^{\dagger}s_1}),
    \end{split}
\end{equation}
\begin{equation}
    \begin{split}
        \dot{ C^c_{s_1^{\dagger}s_2^{\dagger}}} = &-\Gamma_1 C^c_{s_1^{\dagger}s_1^{\dagger}} +\left(-\Gamma_1 -\frac{\kappa_1+\kappa_2}{2} -i(\Delta_1+\Delta_2) \right)C^c_{s_1^{\dagger}s_2^{\dagger}} - \kappa_2 C^c_{s_1^{\dagger}s_2^{\dagger}} C^c_{s_2^{\dagger}s_2} -\kappa_2 C^c_{s_1^{\dagger}s_2}C^c_{s_2^{\dagger}s_2^{\dagger}}
        \\
        & -i\Lambda\left(2C^c_{s_1^{\dagger}s_2^{\dagger}}C^c_{s_2^{\dagger}s_2} + C^c_{s_1^{\dagger}s_2}C^c_{s_2^{\dagger}s_2^{\dagger}} + 2C^c_{s_1^{\dagger}s_2^{\dagger}}\avg{\hat{s}_2}^c \avg{\hat{s}_2^{\dagger}}^c+C^c_{s_1^{\dagger}s_2}\avg{\hat{s}_2^{\dagger}}^{c^2} \right) + ig_2e^{-i\phi_2}C^c_{s_1^{\dagger}s_2} +ig_1e^{-i\phi_1} C^c_{s_2^{\dagger}s_1},
    \end{split}
\end{equation}
\begin{equation}
    \begin{split}
        \dot{ C^c_{s_2^{\dagger}s_2^{\dagger}}} = &-2\Gamma_1 C^c_{s_1^{\dagger}s_2^{\dagger}} + (-\Gamma_1-\kappa_2-i2\Delta_2)C^c_{s_2^{\dagger}s_2^{\dagger}} -2\kappa_2 C^c_{s_2^{\dagger}s_2} C^c_{s_2^{\dagger}s_2^{\dagger}}
        \\
        &-i\Lambda\left(C^c_{s_2^{\dagger}s_2^{\dagger}}+6C^c_{s_2^{\dagger}s_2}C^c_{s_2^{\dagger}s_2^{\dagger}} +4 C^c_{s_2^{\dagger}s_2^{\dagger}}\avg{\hat{s}_2}^c \avg{\hat{s}_2^{\dagger}}^c + \avg{\hat{s}_2^{\dagger}}^{c^2} + 2 C^c_{s_2^{\dagger}s_2}\avg{\hat{s}_2^{\dagger}}^{c^2} \right) +ig_2e^{-i\phi_2}(1+2C^c_{s_2^{\dagger}s_2}).
    \end{split}
\end{equation}
\label{eq-app-steoms}
\end{subequations}
We can also derive the complete set equations of motion for the unconditional dynamics of the cumulants, i.e. noise-averaged equations of motion, as follows:
\begin{subequations}
    \begin{equation}
    \avg{\dot{\hat{s}}_1} = -\frac{\Gamma_1+\kappa_1}{2} \avg{\hat{s}_1} +i\Delta_1 \avg{\hat{s}_1} -ig_1e^{i\phi_1} \avg{\hat{s}_1^{\dagger}}, 
    \label{eq-app-teom-s1}
\end{equation}
\begin{equation}
    \begin{split}
    \avg{\dot{\hat{s}}_2} = & -{\Gamma_1}\avg{\hat{s}_1} + \left(i\Delta_2 -\frac{\kappa_2+\Gamma_1}{2}\right) \avg{\hat{s}_2} +i\Lambda\left(2 C_{s_2^{\dagger}s_2} \avg{\hat{s}_2} + \avg{\hat{s}_2^{\dagger}}\left( \avg{\hat{s}_2}^{2} +  C_{s_2s_2}\right) \right) -ig_2e^{i\phi_2} \avg{\hat{s}_2^{\dagger}} -\sqrt{\kappa_2}\eta_{d,2}e^{i\phi_{d,2}}, 
    \end{split}
    \label{eq-app-teom-s2}
\end{equation}
\begin{equation}
\avg{\dot{\hat{s}}_1^{\dagger}} = -\frac{\Gamma_1+\kappa_1}{2} \avg{\hat{s}_1^{\dagger}} -i\Delta_1 \avg{\hat{s}_1^{\dagger}} +ig_1e^{-i\phi_1} \avg{\hat{s}_1},
\label{eq-app-teom-s1D}
\end{equation}
\begin{equation}
    \begin{split}
   \avg{\dot{\hat{s}}_2^{\dagger}} = & -{\Gamma_1}\avg{\hat{s}_1^{\dagger}} -\left(i\Delta_2 +\frac{\kappa_2+\Gamma_1}{2} \right) \avg{\hat{s}_2^{\dagger}} -i\Lambda\left(2 C_{s_2^{\dagger}s_2} \avg{\hat{s}_2^{\dagger}} + \avg{\hat{s}_2}\left(\avg{\hat{s}_2^{\dagger}}^{2} +C_{s_2^{\dagger}s_2^{\dagger}}\right) \right) +ig_2e^{-i\phi_2} \avg{\hat{s}_2} -\sqrt{\kappa_2}\eta_{d,2}e^{-i\phi_{d,2}}, 
    \end{split}
    \label{eq-app-teom-s2D}
\end{equation}
\begin{equation}
    \begin{split}
        \dot{ C_{s_1s_1}} = (-\Gamma_1-\kappa_1+i2\Delta_1) C_{s_1s_1}   - ig_1e^{i\phi_1}(1+2 C_{s_1^{\dagger}s_1}),
    \end{split}
\end{equation}
\begin{equation}
    \begin{split}
        \dot{ C_{s_1s_2}} = &-\Gamma_1 C_{s_1s_1} + \left(-\Gamma_1 - \frac{\kappa_1+\kappa_2}{2} +i(\Delta_1 + \Delta_2)\right)C_{s_1s_2}
        \\
        &+i\Lambda\left(C_{s_2^{\dagger}s_1}C_{s_2s_2}+ 2C_{s_1s_2}C_{s_2^{\dagger}s_2} + C_{s_2^{\dagger}s_1}\avg{\hat{s}_2}^2 + 2C_{s_1s_2}\avg{\hat{s}_2}\avg{\hat{s}_2^{\dagger}}\right) 
        -ig_1e^{i\phi_1}C_{s_1^{\dagger}s_2} -ig_2e^{i\phi_2}C_{s_2^{\dagger}s_1},
    \end{split}
\end{equation}
\begin{equation}
    \begin{split}
        \dot{ C_{s_2s_2}} = &-2\Gamma_1 C_{s_1s_2} + (-\Gamma_1-\kappa_2 +i2\Delta_2)C_{s_2s_2}
        \\
        &+ i\Lambda\left(C_{s_2s_2} + 6C_{s_2s_2}C_{s_2^{\dagger}s_2} + \avg{\hat{s}_2}^2 + 2C_{s_2^{\dagger}s_2} \avg{\hat{s}_2}^2 + 4C_{s_2s_2} \avg{\hat{s}_2} \avg{\hat{s}_2^{\dagger}} \right) -ig_2e^{i\phi_2}\left(1+2C_{s_2^{\dagger}s_2}\right),  
    \end{split}
\end{equation}
\begin{equation}
    \begin{split}
        \dot{ C_{s_1^{\dagger}s_1}} = (-\Gamma_1-\kappa_1)C_{s_1^{\dagger}s_1}  -i g_1 e^{i\phi_1} C_{s_1^{\dagger}s_1^{\dagger}} + ig_1 e^{-i\phi_1} C_{s_1 s_1},
    \end{split}
\end{equation}
\begin{equation}
    \begin{split}
        \dot{ C_{s_1^{\dagger}s_2}} = & -\Gamma_1 C_{s_1^{\dagger}s_1} + (-\Gamma_1 - \frac{\kappa_1+\kappa_2}{2} + i(-\Delta_1+\Delta_2))C_{s_1^{\dagger}s_2} 
        \\
        &+ i\Lambda \left(2C_{s_1^{\dagger}s_2}C_{s_2^{\dagger}s_2} + C_{s_1^{\dagger}s_2^{\dagger}}C_{s_2 s_2} + C_{s_1^{\dagger}s_2^{\dagger}}\avg{\hat{s}_2}^2 + 2 C_{s_1^{\dagger}s_2}\avg{\hat{s}_2}\avg{\hat{s}_2^{\dagger}}\right) + ig_1e^{-i\phi_1}C_{s_1s_2} -ig_2e^{i\phi_2} C_{s_1^{\dagger}s_2^{\dagger}},
    \end{split}
\end{equation}
\begin{equation}
    \begin{split}
        \dot{ C_{s_2^{\dagger}s_1}} =  &-\Gamma_1 C_{s_1^{\dagger}s_1} + \left(-\Gamma_1 -\frac{\kappa_1+\kappa_2}{2} + i(\Delta_1 - \Delta_2) \right) C_{s_2^{\dagger}s_1}
        \\
        &-i\Lambda\left( 2C_{s_2^{\dagger}s_1}C_{s_2^{\dagger}s_2} +C_{s_1s_2}C_{s_2^{\dagger}s_2^{\dagger}}+ 2C_{s_2^{\dagger}s_1}\avg{\hat{s}_2}\avg{\hat{s}_2^{\dagger}} + C_{s_1s_2} \avg{\hat{s}_2^{\dagger}}^2 \right) + ig_2e^{-i\phi_2} C_{s_1s_2} -ig_1e^{i\phi_1}C_{s_1^{\dagger}s_2^{\dagger}},
    \end{split}
\end{equation}
\begin{equation}
    \begin{split}
        \dot{ C_{s_2^{\dagger}s_2}} = &-\Gamma_1 C_{s_1^{\dagger}s_2} -\Gamma_1 C_{s_2^{\dagger}s_1} + (-\Gamma_1-\kappa_2)C_{s_2^{\dagger}s_2}  
        \\
        &+ i\Lambda\left(C_{s_2^{\dagger}s_2^{\dagger}} \avg{\hat{s}_2}^2 - C_{s_2s_2}\avg{\hat{s}_2^{\dagger}}^2\right) -ig_2e^{i\phi_2} C_{s_2^{\dagger}s_2^{\dagger}}+ig_2e^{-i\phi_2} C_{s_2s_2},
    \end{split}
\end{equation}
\begin{equation}
    \begin{split}
        \dot{ C_{s_1^{\dagger}s_1^{\dagger}}} = (-\Gamma_1-\kappa_1-i2\Delta_1) C_{s_1^{\dagger}s_1^{\dagger}} + ig_1e^{-i\phi_1}(1+2C_{s_1^{\dagger}s_1}),
    \end{split}
\end{equation}
\begin{equation}
    \begin{split}
        \dot{ C_{s_1^{\dagger}s_2^{\dagger}}} = &-\Gamma_1 C_{s_1^{\dagger}s_1^{\dagger}} +\left(-\Gamma_1 -\frac{\kappa_1+\kappa_2}{2} -i(\Delta_1+\Delta_2) \right)C_{s_1^{\dagger}s_2^{\dagger}}
        \\
        & -i\Lambda\left(2C_{s_1^{\dagger}s_2^{\dagger}}C_{s_2^{\dagger}s_2} + C_{s_1^{\dagger}s_2}C_{s_2^{\dagger}s_2^{\dagger}} + 2C_{s_1^{\dagger}s_2^{\dagger}}\avg{\hat{s}_2} \avg{\hat{s}_2^{\dagger}}+C_{s_1^{\dagger}s_2}\avg{\hat{s}_2^{\dagger}}^2 \right) + ig_2e^{-i\phi_2}C_{s_1^{\dagger}s_2} +ig_1e^{-i\phi_1} C_{s_2^{\dagger}s_1},
    \end{split}
\end{equation}
\begin{equation}
    \begin{split}
        \dot{ C_{s_2^{\dagger}s_2^{\dagger}}} = &-2\Gamma_1 C_{s_1^{\dagger}s_2^{\dagger}} + (-\Gamma_1-\kappa_2-i2\Delta_2)C_{s_2^{\dagger}s_2^{\dagger}} 
        \\
        &-i\Lambda\left(C_{s_2^{\dagger}s_2^{\dagger}}+6C_{s_2^{\dagger}s_2}C_{s_2^{\dagger}s_2^{\dagger}} +4 C_{s_2^{\dagger}s_2^{\dagger}}\avg{\hat{s}_2} \avg{\hat{s}_2^{\dagger}} + \avg{\hat{s}_2^{\dagger}}^2 + 2 C_{s_2^{\dagger}s_2}\avg{\hat{s}_2^{\dagger}}^2 \right) +ig_2e^{-i\phi_2}(1+2C_{s_2^{\dagger}s_2}).
    \end{split}
\end{equation}
\label{eq-app-teoms}
\end{subequations}

\section{\label{app-filt}Filtered single-shot measurement mean and variances for linearized systems}

Here we present the calculation of filtered mean and covariance results for a more general system with k modes, starting from the measured trajectories. The mean value of filtered single-shot measurements in the infinite-shot limit is given by:
\begin{align}
    \mu_k(t) = \mathbb{E}\left[ \begin{pmatrix}
        I_k(t)
        \\
        Q_k(t)
    \end{pmatrix} \right] = \frac{1}{\sqrt{2\mathcal{T}}} \int_{t-\mathcal{T}}^{t} d\tau \mathbb{E}\left[
    \begin{pmatrix}
        \mathcal{I}_k(\tau)
        \\
        \mathcal{Q}_k(\tau)
    \end{pmatrix} \right],
    \label{eqn-filt-mean}
\end{align}
with
\begin{subequations}
\begin{align}
    \mathcal{I}_k(t) = \xi_{\mathcal{I}_{k}}(t) + \sqrt{\gamma_H}\left( \left\langle \hat{I}_k(t)\right\rangle +\xi_{\mathcal{I}_{k}}^{qm}(t)\right) +\sqrt{\Bar{n}_{cl}}\xi_{\mathcal{I}_{k}}^{cl}(t),
    \label{eqn-noisy-Itrace}
\end{align}
\begin{align}
\mathcal{Q}_k(t) = \xi_{\mathcal{Q}_{k}}(t) + \sqrt{\gamma_H}\left( \left\langle \hat{Q}_k(t)\right\rangle +\xi_{\mathcal{Q}_{k}}^{qm}(t)\right) +\sqrt{\Bar{n}_{cl}}\xi_{\mathcal{Q}_{k}}^{cl}(t).
\label{eqn-noisy-Qtrace}
\end{align}
\label{eqn-noisy_IQtrace}
\end{subequations}
We should note that  $\xi_{\mathcal{I,Q}_k}(t) = \frac{dW_{I,Q}}{dt}$ where $dW_{I,Q}$ are the Wiener increments that we introduce in the SQME of the system. Secondly, $\xi_{\mathcal{I,Q}_k}^{cl}$ are white noise random variables which represent classical noise contribution to the system after Heisenberg-von Neumann cut. Then we plug in Eqs.~\ref{eqn-noisy_IQtrace} into Eq.~\ref{eqn-filt-mean} and get:
\begin{align}
    \mu_k(t) = \sqrt{\frac{\gamma_H}{2\mathcal{T}}}\int_{t-\mathcal{T}}^{t} d\tau  \begin{pmatrix}
        \left\langle \hat{I}_k(\tau) \right\rangle
        \\
        \left\langle \hat{Q}_k(\tau) \right\rangle
    \end{pmatrix}.
\end{align}
In the steady-state limit ($t\rightarrow \infty$), $\left\langle(\hat{I},\hat{Q})_k(t)\right\rangle$ will settle to a fixed value $\left\langle(\hat{I},\hat{Q})_k\right\rangle$:
\begin{align}
    \mu_k = \lim_{t\rightarrow \infty} \mu_k(t) = \sqrt{\frac{\gamma_{H}\mathcal{T}}{2}} \begin{pmatrix}
        \left\langle \hat{I}_k \right\rangle
        \\
        \left\langle \hat{Q}_k \right\rangle
    \end{pmatrix}.
\end{align}
The covariance matrix for modes j and k of filtered single-shot measurements in the infinite-shot limit is as follows:
\begin{align}
    \Sigma_{jk}(t) &= \frac{1}{2\mathcal{T}} \int_{t-\mathcal{T}}^{t}d\tau \int_{t-\mathcal{T}}^{t}d\tau' \left\{ \mathbb{E}\left[ \begin{pmatrix}
        \mathcal{I}_j(\tau')
        \\
        \mathcal{Q}_j(\tau')
    \end{pmatrix} \begin{pmatrix}
        \mathcal{I}_k(\tau)
        \\
        \mathcal{Q}_k(\tau)
    \end{pmatrix}^T\right] - \mathbb{E}\left[\begin{pmatrix}
        \mathcal{I}_j(\tau')
        \\
        \mathcal{Q}_j(\tau')
 \end{pmatrix}\right] \mathbb{E}\left[\begin{pmatrix}
        \mathcal{I}_k(\tau)
        \\
        \mathcal{Q}_k(\tau)
 \end{pmatrix}\right]^T \right\} 
 \\ &= \frac{1}{2\mathcal{T}} \int_{t-\mathcal{T}}^{t}d\tau \int_{t-\mathcal{T}}^{t}d\tau'  \hat{s}_{jk}(\tau',\tau), 
 \label{eqn-FiltCov}
\end{align}
where
\begin{align}
    \hat{s}_{jk}(\tau',\tau) = \delta_{jk}\delta(\tau'-\tau)(1+\Bar{n}_{cl})\mathbb{I}_2 + M_j C(\tau',\tau)M_k^T,
    \label{eqn-cov-ipop}
\end{align}
with
\begin{align}
    \left[C(\tau',\tau)\right]_{ij} = \left \langle :z_i(\tau')z_j(\tau):\right\rangle - \left\langle z_i(\tau')\right\rangle\left\langle z_j(\tau)\right\rangle,
    \label{eqn-covmat-intracav}
\end{align}
\begin{align}
    \vec{z} = \begin{pmatrix}
        b_1 & b_1^{\dagger} & ... & b_k & b_k^{\dagger} & ... & b_K & b_K^{\dagger}
    \end{pmatrix}^T,
    \label{eqn-system-fieldops}
\end{align}
\begin{align}
    M_k = \sqrt{\gamma_H} \begin{pmatrix}
        0 & ... & \underbrace{U}_{k^{th} elem} & ... & 0
    \end{pmatrix},
    \label{eqn-ca-to-quadrature-transformation}
\end{align}
\begin{align}
    U = \frac{1}{\sqrt{2}}\begin{pmatrix}
        1&1\\-i&i
    \end{pmatrix},
    \label{eqn-unitary-quadmat}
\end{align}
and $b_k^{(\dagger)}$ are creation and annihilation operators for intracavity modes, K is the total number of modes in the system. Plugging these in, we get:
\begin{equation}
    \begin{split}
    \Sigma_{jk}(t) = 
    \frac{1}{2}\delta_{jk}(1+\Bar{n}_{cl}) \mathbb{I}_2+ \frac{1}{2\mathcal{T}}  \int_{t-\mathcal{T}}^{t}d\tau \int_{t-\mathcal{T}}^{t}d\tau' M_j C(\tau',\tau)M_k^T.
    \end{split}
\end{equation}
For linear and linearized systems the following statements hold:
\begin{align}
    C(\tau+\theta,\tau) &= e^{J\theta}C(\tau),
    \\ C(\tau-\theta,\tau) &= C(\tau-\theta)e^{J^{T}}.
\end{align}
We can break down the double integral into two as:

\begin{align}
    \int_{t-\mathcal{T}}^{t} d\tau \int_{t-\mathcal{T}}^{t} d\tau'  = \int_{t-\mathcal{T}}^{t} d\tau \int_{t-\mathcal{T}}^{\tau} d\tau' + \int_{t-\mathcal{T}}^{t} d\tau \int_{\tau}^{t} d\tau' .
\end{align}
See in the first integral $\tau'<\tau$ so we can define $\tau'=\tau-\theta$ and in the second integral $\tau'> \tau$ so we can define $\tau'=\tau+\theta$ and rewrite the integrals in terms of $\tau$ and $\theta$.
\begin{equation}
    \begin{split}
        \int_{t-\mathcal{T}}^{t}d\tau \int_{t-\mathcal{T}}^{t}d\tau' M_j C(\tau',\tau)M_k^T  &= \int_{t-\mathcal{T}}^{t}d\tau \int_{0}^{\tau-(t-\mathcal{T})}d\theta M_j C(\tau-\theta,\tau)M_k^T 
        + \int_{t-\mathcal{T}}^{t}d\tau \int_{0}^{t-\tau} d\theta M_j C(\tau+\theta,\tau)M_k^T
        \\
        &= \int_{t-\mathcal{T}}^{t}d\tau \int_{0}^{\tau-(t-\mathcal{T})}d\theta M_j C(\tau-\theta)e^{J^T\theta}M_k^T 
        + \int_{t-\mathcal{T}}^{t}d\tau \int_{0}^{t-\tau} d\theta M_j e^{J\theta}C(\tau)M_k^T.
    \end{split}
\end{equation}
Then, we go to the steady-state limit as $t\rightarrow \infty$, $C(t)=C(t-\theta)=C$
\begin{equation}
    \begin{split}
        \int_{t-\mathcal{T}}^{t}d\tau \int_{t-\mathcal{T}}^{t}d\tau' M_j C(\tau',\tau)M_k^T  
        &= \int_{t-\mathcal{T}}^{t}d\tau \int_{0}^{\tau-(t-\mathcal{T})}d\theta M_j C e^{J^T\theta}M_k^T 
        + \int_{t-\mathcal{T}}^{t}d\tau \int_{0}^{t-\tau} d\theta M_j e^{J\theta} C M_k^T.
    \end{split}
\end{equation}
We use the diagonal representation of matrix $J$ to continue our calculation, $J = PNP^{-1}$. With this representation then we rewrite $e^{J\theta} = Pe^{N\theta}P^{-1}$ where $N$ is a diagonal matrix
\begin{equation}
    \begin{split}
        \int_{t-\mathcal{T}}^{t} d\tau \int_{0}^{t-\tau} d\theta e^{J\theta} &= P \left[\int_{t-\mathcal{T}}^{t} d\tau \int_{0}^{t-\tau} d\theta e^{N\theta} \right]P^{-1}
        \\
        &= PN^{-1}\int_{t-\mathcal{T}}^{t} d\tau \left[e^{N(t-\tau)}-\mathbb{I}\right] P^{-1}
        \\
        &=PN^{-2}\left[e^{N\mathcal{T}}-\mathbb{I}\right]P^{-1} - \mathcal{T}J^{-1}
        \\
        &= J^{-2 \left[e^{J\mathcal{T}}-\mathbb{I}\right]} - \mathcal{T}J^{-1},
    \end{split}
\end{equation}
where we used $J^{n} = PN^nP^{-1}$ for all $n \in \mathbb{Z}$. Next, we can use $J^{T^n} = P^{T^-1}N^nP^T$.
\begin{equation}
    \begin{split}
    \int_{t-\mathcal{T}}^{t} d\tau \int_{0}^{\tau-(t-\mathcal{T})} d\theta e^{J^T\theta} &= (P^T)^{-1} \int_{t-\mathcal{T}}^{t} d\tau \int_{0}^{\tau-(t-\mathcal{T})} d\theta e^{N\theta} P^T
    \\
    &= (P^T)^{-1} N^{-1}\int_{t-\mathcal{T}}^{t} d\tau \left[e^{N(\tau-((t-\mathcal{T})} - \mathbb{I}\right] P^T
    \\
    &= (P^T)^{-1} N^{-2}\left[e^{N\mathcal{T}} -\mathbb{I}\right] P^T - \mathcal{T}(J^T)^{-1}
    \\
    &= (J^T)^{-2}\left[e^{J^T \mathcal{T}} -\mathbb{I} \right] - \mathcal{T}(J^T)^{-1}.
    \end{split}
\end{equation}
Plugging these in, we get:
\begin{align}
    \Sigma_{jk} = \frac{1}{2}\delta_{jk}(1+\Bar{n}_{cl})\mathbb{I}_2 -\frac{1}{2}M_j \left[ CJ^{T^{-1}}  +J^{-1}  C 
        + \frac{1}{\mathcal{T}}\left(CJ^{T^{-2}}\left(\mathbb{I}-e^{J^T\mathcal{T}}\right) + J^{-2} \left(\mathbb{I}-e^{J \mathcal{T}}\right) C\right) \right] M_k^T. 
\end{align}
In the limit where the integration time is very long ($\mathcal{T} \rightarrow \infty$), we end up with the following expression for the covariance matrix (as the eigenvalues of $J$ have all real negative parts)
\begin{equation}
\begin{split}
    \lim_{\mathcal{T} \rightarrow \infty} \Sigma_{jk} &= \frac{1}{2}\delta_{jk}(1+\Bar{n}_{cl})\mathbb{I}_2 -\frac{1}{2}M_j \left[ CJ^{T^{-1}}  +J^{-1}  C 
        + \frac{1}{\mathcal{T}}\left(CJ^{T^{-2}} + J^{-2}   C\right) \right] M_k^T
    \\
    &= \frac{1}{2}\delta_{jk}(1+\Bar{n}_{cl})\mathbb{I}_2 -\frac{1}{2}M_j \left[ CJ^{T^{-1}}  +J^{-1}  C \right] M_k^T + \mathcal{O}\left(\frac{1}{\mathcal{T}}\right).
    \end{split}
\end{equation}

\section{\label{app-pert}A non-traditional perturbative approach to evaluate cumulants in weak nonlinearity limit}

In this section, we present a perturbative treatment on the TEOMs. The correct ansatz of the perturbative expansion depends only on the type of nonlinearity in the system, so in order to find the correct expansion coefficients we start with the TEOMs of a single Kerr oscillator, which is much simpler than the system that we work with but still gives us the correct expansion coefficients:
\begin{subequations}
\begin{align}
        \dot{s} &= \left(i\Delta - \kappa/2\right) s -\sqrt{\kappa}\eta e^{i\phi_d} + i\Lambda \left(s^{\dagger} s s + C_{ss} s^{\dagger} + 2 C_{s^{\dagger}s} s \right),
        \\
        \dot{C}_{ss} &= \left(i2\Delta - \kappa \right) C_{ss} + i\Lambda \left(C_{ss} \left(1+6C_{s^{\dagger}s} + 4 s s^{\dagger}\right) +  s^2 \left(1+2C_{s^{\dagger}s}\right) \right),
        \\
        \dot{C}_{s^{\dagger}s} &= -\kappa C_{s^{\dagger}s} + i\Lambda \left(C_{s^{\dagger}s^{\dagger}}s^2 - C_{ss} s^{\dagger^2} \right).
\end{align}
\label{eqn:Kerr-TEOMs}
\end{subequations}
Then, we expand the first order cumulants with the following ansatz:
\begin{align}
    s = \Lambda^{\alpha} s^{(0)} + \Lambda^{\beta} s^{(1)},
    \label{eq-pert-s}
\end{align}
with $\beta > \alpha$, which guarantees that $s^{(0)}$ is the leading order term. Then we expand the second order cumulants with the following ansatz:
\begin{align}
    C = \Lambda^p C^{(0)} + \Lambda^{q} C^{(1)},
    \label{eq-pert-C}
\end{align}
similarly with $q > p$, which guarantees that $C_{ss}^{(0)}$ is the leading order term. Now, we plug in these expressions in Eq.~(\ref{eqn:Kerr-TEOMs}):
\begin{equation}
    \begin{split}
        &\Lambda^{\alpha}\dot{s}^{(0)} + \Lambda^{\beta} \dot{s}^{(1)} = \left(i\Delta-\kappa/2\right) \left( \Lambda^{\alpha} s^{(0)} + \Lambda^{\beta} s^{(1)} \right) - \sqrt{\kappa}\eta e^{i\phi_d} 
        \\
        &+ i\left[\Lambda^{1+3\alpha}s^{(0)^2}s^{\dagger^{(0)}} + \Lambda^{1+2\alpha+\beta} \left(2 s^{\dagger^{(0)}} s^{(0)} s^{(1)} + s^{(0)^2}  s^{{\dagger}^{(1)}}\right)+ \Lambda^{1+\alpha+2\beta} \left(2 s^{(0)} s^{(1)} s^{\dagger^{(1)}} + s^{\dagger^{(0)}} s^{(1)^2} \right) +\Lambda^{1+3\beta} s^{(1)^2} s^{\dagger^{(1)}} \right]
        \\
        &+i \left[\Lambda^{1+p+\alpha}\left(C_{ss}^{(0)}s^{\dagger^{(0)}} + 2C_{s^{\dagger}s}^{(0)}s^{(0)}\right) + \Lambda^{1+p+\beta}\left(C_{ss}^{(0)}s^{\dagger^{(1)}} + 2C_{s^{\dagger}s}^{(0)}s^{(1)}\right) + \Lambda^{1+q+\alpha}\left(C_{ss}^{(1)}s^{\dagger^{(0)}} + 2C_{s^{\dagger}s}^{(1)}s^{(0)}\right)  \right.\\
         &+ \left. \Lambda^{1+q+\beta}\left(C_{ss}^{(1)}s^{\dagger^{(1)}}+ 2C_{s^{\dagger}s}^{(1)}s^{(1)}\right) \right].
    \end{split}
\end{equation}
The lowest-order term for this equation has $\Lambda^{\alpha}$ coefficient and the terms on this order only include leading-order terms for first-order cumulants and a drive term and is equivalent to the normalized equation of motion of the Kerr oscillator in the classical limit:
\begin{equation}
    \begin{split}
        \Lambda^{\alpha} \left[\dot{s}^{(0)} = \left(i\Delta-\kappa/2\right)s^{(0)} - \sqrt{\kappa}\eta\Lambda^{-\alpha}e^{i\phi_d} +  i\Lambda^{1+2\alpha} s^{(0)^2}s^{\dagger^{(0)}}\right]. 
    \end{split}
\end{equation}
In order to have all the prefactors matching, we need $1+2\alpha = 0$, which fixes prefactor of the leading- order mean value: 
\begin{align}
    \alpha = -1/2.
\end{align}
Notice that, in this case, the drive is normalized by the nonlinearity $\eta' = \eta \sqrt{\Lambda}$. This means that for a fixed value of $\eta'$, one ends up with identical dynamics even if the nonlinearity (or drive) is changed. We should also notice that this coefficient will always be decided by the type of nonlinearity one has in the system dynamics. 

Moving forward, by plugging in $\alpha=-1/2$, we write down the second largest terms in this expansion, which are scaled with $\Lambda^{\beta}$:
\begin{equation}
\begin{split}
    \Lambda^{\beta} \left[ \dot{s}^{(1)} = \left(i\Delta-\kappa/2\right) s^{(1)}  + i2s^{\dagger^{(0}}s^{(0)}s^{(1)} +  i s^{(0)^2}s^{\dagger^{(1)}} + i\Lambda^{\frac{1}{2}+p-\beta} \left( C_{ss}^{(0)}s^{\dagger^{(0)}} + 2C_{s^\dagger s}^{(0)}s^{(0)}\right) \right].
\end{split}
\end{equation}
Here, we keep the lowest-order term coming from the second-order cumulant contribution, which forces $\frac{1}{2}+p-\beta = 0$. We have a new condition for the perturbative coefficients:
\begin{align}
    \beta = p+\frac{1}{2}.
\end{align}
Now, we plug in our perturbative ansatz into the second-order cumulant equation of motion:
\begin{equation}
    \begin{split}
        \Lambda^{p} \dot{C}_{ss}^{(0)} + \Lambda^q \dot{C}_{ss}^{(1)} = &\left(i2\Delta-\kappa\right) \left(\Lambda^{p} {C}_{ss}^{(0)} + \Lambda^q {C}_{ss}^{(1)}\right) + i\left[\Lambda^{p+1}C_{ss}^{(0)} + \Lambda^{q+1}C_{ss}(1) + 6\Lambda^{1+2p} C_{ss}^{(0)} C_{s^\dagger s}^{(0)} \right.
        \\
        &+ \left. 6\Lambda^{1+p+q} \left(C_{ss}^{(0)}C_{s^{\dagger}s}^{(1)} + C_{s^\dagger s}^{(0)}C_{ss}^{(1)}\right) + 6\Lambda^{1+2q} C_{ss}^{(1)}C_{s^{\dagger}s}^{(1)} + 4\Lambda^{p} C_{ss}^{(0)}s^{(0)}s^{\dagger^{(0)}} \right.
        \\
        &+ \left. 4 \Lambda^{\frac{1}{2}+p+\beta} \left(C_{ss}^{(0)}\left(s^{(0)}s^{\dagger^{(1)}} + s^{\dagger^{(0)}}s^{(1)}\right)\right) + 4\Lambda^{1+p+2\beta}C_{ss}^{(0)}s^{(1)}s^{\dagger^{}(1)} + 4\Lambda^{q}C_{ss}^{(1)}s^{(0)}s^{\dagger^{(0)}} \right. 
        \\
        &+ \left. 4\Lambda^{\frac{1}{2}+q+\beta} \left(C_{ss}^{(1)}\left(s^{(0)}s^{\dagger^{(1)}} + s^{\dagger^{(0)}}s^{(1)}\right)\right) + 4\Lambda^{1+q+2\beta} C_{ss}^{(1)} s^{(1)} s^{\dagger^{(1)}}
        \right] + i\left[s^{(0)^2}
        +2\Lambda^{\beta+\frac{1}{2}}s^{(0)}s^{(1)}\right. 
        \\
        &+ \left. \Lambda^{1+2\beta}s^{(1)^2} + \Lambda^{p} 2C_{ss}^{(0)}s^{(0)^2} + \Lambda^{\frac{1}{2}+p+\beta} 2 C_{ss}^{(0)}s^{(0)}s^{(1)} + \Lambda^{1+p+2\beta}2C_{ss}^{(0)}s^{(1)^2} \right. 
        \\
        &+ \left.  \Lambda^{q} 2C_{ss}^{(1)}s^{(0)^2} + \Lambda^{\frac{1}{2}+q+\beta} 2 C_{ss}^{(1)}s^{(0)}s^{(1)} + \Lambda^{1+q+2\beta}2C_{ss}^{(1)}s^{(1)^2}
        \right].
    \end{split}
\end{equation}
Ww write down the lowest-order components:
\begin{equation}
    \begin{split}
        \Lambda^p\left[ \dot{C}_{ss}^{(0)} = \left(i2\Delta - \kappa\right)C_{ss}^{(0)} + 4C_{ss}^{(0)}s^{(0)}s^{\dagger^{(0)}} +\Lambda^{-p} s^{(0)^2} + 2C_{ss}^{(0)} s^{(0)^2} \right].
    \end{split}
\end{equation}
This equation fixes the value of $p$ and $\beta$, by forcing $-p=0$:
\begin{align}
    p&=0,
    \\
    \beta &= \frac{1}{2}. 
\end{align}
Finally, to find $q$ we write down the second lowest-order terms in this equation:
\begin{equation}
    \begin{split}
        \Lambda^{q} \Big[ &\dot{C}_{ss}^{(1)} = \left(i2\Delta - \kappa\right) C_{ss}^{(1)} + i\Lambda^{1-q} 6 C_{ss}^{(0)} C_{s^{\dagger}s}^{(0)} +  i\Lambda^{1-q} 4C_{ss}^{(0)} \left(s^{(0)} s^{\dagger^{(1)}} + s^{\dagger^{(0)}} s^{(1)}\right) + i 4 s^{(0)} s^{\dagger^{(0)}} C_{ss}^{(1)}  
        \\
         &+ i\Lambda^{1-q} 2 s^{(0)} s^{(1)} + i\Lambda^{1-q} 4 C_{ss}^{(0)}  s^{(0)} s^{(1)} + i 2 s^{(0)^2} C_{ss}^{(1)}
        \Big].
    \end{split}
\end{equation}
This finally fixes $q$ by forcing $1-q=0$:
\begin{align}
    q=1.
\end{align}
We plug these coefficients into Eq.~(\ref{eq-pert-s},\ref{eq-pert-C}), and then use this ansatz to evaluate the TEOMs for the system in the main text.

Notice that this perturbative expansion gives us hierarchical equations of motion starting from determining the zeroth-order solution of the mean value, which is equivalent to the equation of motion of the system in the classical limit, where the mode amplitude is normalized by the nonlinearity. Also, notice that this is the only nonlinear equation that needs to be solved. Compared to the traditional perturbative expansions, this is a major difference, as this derivation takes into account the effect of nonlinearity in the zeroth-order terms for mode means.

The solution of this equation is then used to determine the linear equations of motion for the zeroth-order contribution to the second-order cumulants. This solution, along with the previous solution, is then used to determine the linear equations of motion for the first-order contribution to the mean. All of these solutions are then used to determine the linear equations of motion for the first-order contribution to the second-order cumulants. This is where we stopped our expansion but one can start with an ansatz including higher-order terms to extend this approximation. The same procedure needs to be done to find the coefficients for the higher-order perturbative terms and the same pattern would apply to those higher-order terms in the same hierarchical fashion.

In the main text result for the performance metrics (Eq.~(\ref{eq-ms-pert},\ref{eq-ms-pert-J} and \ref{eq-ms-pert-V}), second-order cumulants are approximated by their zeroth-order contribution only ($C \approx C^{(0)}$), and we denote $s^{(0)}$ as $\bar{s}$ and $C^{(0)}$ as $\bar{C}$ in order to avoid confusion with the superscript denoting the input state type.

\bibliographystyle{apsrev4-2}
\end{document}